\documentclass[smallextended,numbook,runningheads]{svjour3}
\smartqed  
\usepackage{graphicx}
\usepackage{amsmath}
\usepackage{epstopdf} 
\usepackage{amssymb}
\usepackage{dsfont}
\usepackage[utf8]{inputenc}
\usepackage[usenames, svgnames]{xcolor}
\usepackage{varwidth}
\usepackage{caption}
\usepackage{hyperref}
\usepackage{ulem}
\usepackage{adjustbox}
\usepackage[ruled]{algorithm}            
\usepackage{algorithmicx} 
\usepackage{algpseudocode}
\usepackage{subcaption}

\makeatletter
\@addtoreset{algorithm}{section}
\makeatother

\algrenewcommand\algorithmicindent{1.0em}
\algrenewcommand{\algorithmicrequire}{\tiny \textbf{\tiny{Input:}}}
\algrenewcommand{\algorithmicensure}{\small{ \textbf{Output:}}}
\algrenewcommand{\alglinenumber}[1]{\color{red!80!blue}\footnotesize#1:}
\usepackage{calc} 
\algrenewcommand\algorithmicfunction{\widthof{\textbf{\small{Function}}}}
\algrenewcommand{\algorithmiccomment}[1]{{\color{Green}\hfill\footnotesize $\vartriangleright$ #1}} 
\algnewcommand\algorithmicto{\textbf{to}}
 \algrenewtext{For}[3]%
   {\algorithmicfor\ $#1 \gets #2$ \algorithmicto\ $#3$ \algorithmicdo}
\makeatletter
\renewcommand{\ALG@beginalgorithmic}{\small}

\let\OldStatex\Statex
\renewcommand{\Statex}[1][3]{%
  \setlength\@tempdima{\algorithmicindent}%
  \OldStatex\hskip\dimexpr#1\@tempdima\relax}
\newcommand{\StatexIndent}[1][3]{%
  \setlength\@tempdima{\algorithmicindent}%
  \Statex\hskip\dimexpr#1\@tempdima\relax}
\makeatother

\newif\ifstartedinmathmode
\newcommand{\FName}[1]{{\footnotesize \color{DarkRed} \ifstartedinmathmode\mbox{\textsc{#1}}\else\textsc{#1}\fi }}
\newcommand{\FNameDef}[1]{{\footnotesize {\color{red} \ifstartedinmathmode\mbox{\textsc{#1}}\else\textsc{#1}\fi }}}
\newcommand{\FNameStd}[1]{{\footnotesize {\color{blue} \ifstartedinmathmode\mbox{\textsc{#1}}\else\textsc{#1}\fi }}}
\newcommand{\Var}[1]{ {\ifstartedinmathmode\mbox{\texttt{#1}}\else\texttt{#1}\fi }}

\newcommand{\Input}[2]{
\begin{minipage}{#1}
{\footnotesize \textbf{Input :}\\
#2} \\
\end{minipage}\\
}
\newcommand{\Output}[2]{
\begin{minipage}{#1}
{\footnotesize \textbf{Output :}\\
#2}
\end{minipage}\\
}

\newcommand{\OptVec}{OptV}
\newcommand{\OptVecS}{OptVS}
\newcommand{\MultVec}{\vecb{.*}}
\newcommand{\DivVec}{\,\vecb{./}}
\newcommand{\PlusVec}{\vecb{+}}

\newcommand{\mi}[1]{\vecb{#1}} 
\newcommand{\XPk}[1]{{X_h^{#1}}} 
\newcommand{\Pk}[1]{{\mathbb{P}_{#1}}} 
\newcommand{\simplex}{K} 
\newcommand{\DOM}{\Omega}
\newcommand{\DOMH}{{\DOM_h}}
\newcommand{\Th}{{\mathcal{T}_h}}
\newcommand{\R}{\mathbb{R}}
\newcommand{\N}{\mathbb{N}}
\newcommand{\ENS}[2]{\left\{ {{#1},\hdots,{#2}}\right\}}
\newcommand{\DOT}[2]{\left\langle #1,#2 \right\rangle}
\newcommand{\GRAD}{\mathop{\rm \nabla}\nolimits}
\newcommand{\DP}[2]{\frac{\partial #1}{\partial #2}}
\newcommand{\bO}[1]{\mathcal{O}\left({#1}\right)}
\newcommand{\MAT}[1]{\mathbb{#1}}
\newcommand{\vecb}[1]{\pmb{#1}}
\newcommand{\VEC}[1]{\vecb{#1}}          
\newcommand{\LInfD}[1]{{L^\infty}({#1})}
\newcommand{\il}{{\rm \alpha}}
\newcommand{\jl}{{\rm \beta}}
\newcommand{\kl}{{\rm \gamma}}
\newcommand{\al}{\mathop{\rm \nu}\nolimits}
\newcommand{\q}{{\rm q}}
\newcommand{\nq}{{\mathop{\rm n_q}\nolimits}}
\newcommand{\me}{\mathop{\rm me}\nolimits}
\newcommand{\nme}{{\mathop{\rm n_{me}}\nolimits}}
\newcommand{\areas}{{\mathop{\rm vols}\nolimits}}
\newcommand{\volumes}{{\mathop{\rm vols}\nolimits}}
\newcommand{\dd}{{\rm d}}
\newcommand{\Kg}{\MAT{K}_g}
\newcommand{\Ig}{\MAT{I}_g}
\newcommand{\Jg}{\MAT{J}_g}
\newcommand{\SMIE}{\mathcal{S}} 
\newcommand{\PihK}[1]{\mathop{\pi_\simplex^{#1}}} 
\newcommand{\BaCo}{\lambda} 
\newcommand{\BasisFunc}{\varphi}
\newcommand{\PkBasisFunc}{\BasisFunc}
\newcommand{\BaryCoor}{\lambda} 
\newcommand{\BaryCoorVec}{\vecb{\lambda}} 
\newcommand{\ndfe}{{\mathop{\rm n_{dfe}}\nolimits}}
\newcommand{\ndfes}{{\mathop{\rm n^2_{dfe}}\nolimits}}
\newcommand{\ndf}{{\mathop{\rm n_{dof}}\nolimits}}
\newcommand{\ndof}{{\mathop{\rm n_{dof}}\nolimits}}
\newcommand{\ndofs}{{\mathop{\rm n^2_{dof}}\nolimits}}
\newcommand{\BasisFuncTwoD}{\vecb{\psi}}    
\newcommand{\Stiff}{\mathbb{S}}
\newcommand{\StiffElem}{\mathbb{S}^e}
\newcommand{\StiffElas}{\mathbb{K}}
\newcommand{\StiffElasElem}{\mathbb{K}^e}
\newcommand{\MassW}[1]{\mathbb{M}^{[#1]}}
\newcommand{\MassWElem}[1]{\mathbb{M}^{[#1],e}}
\newcommand{\Mass}{\mathbb{M}}

\newcommand{\simplices}{simplices }
\newcommand{\Odv}{\underline{\vecb{\epsilon}}}  
\newcommand{\Ocv}{\underline{\vecb{\sigma}}}
\newcommand{\CC}{C\nolinebreak[4]\hspace{-.05em}\raisebox{.4ex}{\tiny\bf ++}}

\newcommand{\MatSet}{\mathcal{M}}
\newcommand{\MS}[2]{\MatSet_{#1}(#2)}

\newcommand{\foncdefsmall}[3]{%
{#1}  : {#2} \longrightarrow  {#3}
}

\newcommand{\zoom}{0.344}

\graphicspath{{./images/}{./}}
\DeclareGraphicsExtensions{.eps}
\newcommand{\imageps}[2]{
  \begin{center}
    \hspace{-0.5cm}
    \includegraphics*[scale={#2}]{#1}
  \end{center}}
\newcommand{\dblimageps}[4]{
  \begin{center}
    \hspace{-0.5cm}
    \includegraphics*[scale={#2}]{#1}
    \includegraphics*[scale={#4}]{#3}
  \end{center}}

\newcommand{\BenchFigureNew}[6]{
\begin{figure}[#6]
  \centering
\dblimageps{#1}{\zoom}
{#2}{\zoom}
\vspace*{-5mm}
\imageps{#3}{\zoom}
\vspace*{-5mm}
\caption{#4 \label{#5}}
\end{figure}
}

\journalname{BIT}
\begin{document}

\title{An efficient way to assemble finite element matrices in vector languages
\thanks{This work was partially funded by GNR MoMaS,
   CoCOA LEFE project, ANR~DEDALES and MathSTIC (University Paris 13)
}
}
\titlerunning{Efficient finite element assembly algorithms in vector languages}

\author{Fran\c{c}ois Cuvelier \and Caroline Japhet \and Gilles~Scarella}

\institute{F. Cuvelier  \and G. Scarella \at
           Universit\'e Paris 13, Sorbonne Paris Cité, LAGA, CNRS UMR 7539,
           99 Avenue J-B Clément, F-93430 Villetaneuse, France,
           \email{cuvelier@math.univ-paris13.fr, scarella@math.univ-paris13.fr}
           \and
           C. Japhet \at             Universit\'e Paris 13, Sorbonne Paris Cité, LAGA, CNRS UMR 7539,
           99 Avenue J-B Clément, F-93430 Villetaneuse, France. INRIA Paris-Rocquencourt,
           BP 105, F-78153 Le Chesnay, France,  \email{japhet@math.univ-paris13.fr}}

\date{Received: date / Accepted: date}
\reversemarginpar 
\maketitle

\begin{abstract}
Efficient Matlab codes in 2D and 3D have been proposed recently to assemble finite element matrices.
In this paper we present simple, compact and efficient vectorized algorithms, which are variants of these
codes, in arbitrary dimension,
without the use of any lower level language. They can be easily implemented in many
vector languages
(e.g. Matlab, Octave, Python, Scilab, R, Julia, \CC \! \! with STL,...).
The principle of these techniques is general, we
present it for the assembly of several finite element matrices
in arbitrary dimension, in the $\Pk{1}$ finite element case.
We also provide an extension of the algorithms
to the case of a system of PDE's.
Then we give an extension to piecewise polynomials of higher order.
We compare numerically the performance of these algorithms in Matlab, Octave and Python,
with that in FreeFEM++ and in a compiled language such as~C.
Examples show that, unlike what is commonly believed,  the performance
is not radically worse than that of C :
in the best/worst cases, selected vector languages are
respectively 2.3/3.5 and 2.9/4.1 times slower than~C in the scalar and vector cases.
We also present numerical results which illustrate the computational costs of these algorithms compared
to standard algorithms and to other recent ones.
\end{abstract}

\section{Introduction}
Vector languages\footnotemark\footnotetext{which contain usual element-wise operators and functions on
multidimensional arrays} such as Matlab~\cite{Matlab:2014},
GNU Octave~\cite{octave:2014}, Python~\cite{Python}, R~\cite{R:2015}, Scilab~\cite{scilab:2015},  Julia~\cite{julia:2015}, \CC \! with STL,..., are very widely used for scientific computing
(see for example~\cite{Chen:iFEM:2013,Fenics:2012,Hesthaven:NDG:2008,Langtangen:OEP:2008,Quarteroni:SCM:2013})
and there is significant interest in programming techniques in these languages for two reasons.
The first concerns how to make clear, compact code to ease implementation and understanding, which is important for
teaching and rapid-prototyping in research and industry.
The second concerns how to make this compact code fast enough for realistic simulations.

On the other hand, in finite element
  simulations~\cite{Chen:FEM:2005,Ciarlet:TFE:2002,Johnson:NSP:2009,Quarteroni:NAP:2008,Thomee:GFE:1997},
  the need for efficient algorithms for assembling the matrices
may be crucial, especially when the matrices may need to be assembled several times. This is the case for example
when simulating time-dependent problems with explicit or implicit schemes with
time-dependent coefficients (e.g. in ocean-atmosphere coupling or porous medium applications).
Other examples are computations with a posteriori estimates when one needs to reassemble the matrix equation
on a finer mesh, or in the context of eigenvalue problems where assembling the matrix may be costly.
In any event, assembly remains a critical part of code optimization since solution of
linear systems, which asymptotically dominates in large-scale computing, could be done with the
linear solvers of the different vector languages.

In a vector language, the inclusion of loops is a critical performance degrading aspect
and removing them is known as a vectorization. In finite element programming,
the classical finite element assembly is based on a loop over the elements
(see for example~\cite{Lucquin:ISC:1998}).
In~\cite{Davis:DMS:2006}
T. Davis describes different assembly techniques applied to random matrices of finite element type.
A first vectorization technique is proposed in~\cite{Davis:DMS:2006}.
Other more efficient algorithms in Matlab have been proposed recently
in~\cite{Anjam:FMA:2014,Chen:PFE:2011,Chen:iFEM:2013,Dabrowski:MIL:2008,Funken:2011,Hannukainen:IFE:2012,Koko:SP:2007,Rahman:FMA:2011}.

In this paper we describe vectorized algorithms, which are variants of the codes
in~\cite{Chen:PFE:2011,Chen:iFEM:2013,Funken:2011,Koko:SP:2007}, extended to
arbitrary dimension $d \ge 1$,
  for assembling large sparse matrices in
  finite element computations.
A particular strength of these algorithms is that they make using, reading and extending the codes easier
while achieving performance close to that of~C.

The aim of this article is the quantitative studies for illustrating the efficiency of the vector languages and the various speed-up of the algorithms, relatively to each other, to~C and to FreeFem++~\cite{FF}.
We also propose a vectorized algorithm in arbitrary dimension which is easily transposable to matrices arising from
PDE's such as (see~\cite{Quarteroni:NAP:2008})
  \begin{equation}\label{eq:EDP}
  -\nabla \cdot\left(\mathbb{A}\nabla u\right)
+ \nabla \cdot\left(\pmb{b}u\right) + \pmb{c} \cdot \nabla u  +a_0 u =f \quad \mbox{in } \DOM,
\end{equation}
where $\DOM$ is a bounded domain of $\R^d$ ($d\geq 1$), $\mathbb{A}\in (L^\infty(\DOM))^{d\times d},$ $\pmb{b}\in(L^\infty(\DOM))^{d},$
$\pmb{c}\in(L^\infty(\DOM))^{d},$ $a_0 \in L^\infty(\DOM)$ and $f \in L^2(\DOM)$ are given functions.
The description of the vectorized algorithm is done in three steps:
we recall (non-vectorized) versions called \texttt{base} and \texttt{OptV1}. The latter requires sparse matrix tools found in most of the languages used for computational science and
engineering. Then we give vectorized algorithms which are much faster: \texttt{OptV2} (memory consuming), \texttt{\OptVec} (less memory consuming) and \texttt{\OptVecS} (a symmetrized version of \texttt{\OptVec}).
These algorithms have been tested for several matrices
  (e.g. weighted mass, stiffness and elastic stiffness matrices) and in different languages.
We also provide an extension to the vector case in arbitrary dimension,
where the algorithm is applied to the elastic stiffness matrix with variable coefficients, in 2D and 3D.

For space considerations, we restrict ourselves in this paper to~$\Pk{1}$ Lagrange finite elements.
However, in the appendix we show that with slight modification, the algorithm is valid for piecewise polynomials
of higher order.

These algorithms can be efficiently implemented in many languages if the
language has a sparse matrix implementation. For the \texttt{OptV1}, \texttt{OptV2}, \texttt{\OptVec} and
  \texttt{\OptVecS} versions,
a particular sparse matrix constructor is also needed (see Section~\ref{sec:ClassAssemb})
and these versions require that the language supports element-wise array operations.
Examples of languages for which we obtained an efficient implementation of these algorithms are
\begin{itemize}
\item[$\bullet$] Matlab,
\item[$\bullet$] Octave,
\item[$\bullet$] Python with \textit{NumPy} and \textit{SciPy} modules,
\item[$\bullet$] Scilab,
\item[$\bullet$] \textit{Thrust} and \textit{Cusp}, C++ libraries for CUDA
\end{itemize}
This paper is organized as follows: in Section~\ref{sec:Notations} we define two examples
  of finite element matrices. Then we introduce the notation associated to the mesh
  and to the algorithmic language used in this article.
In Section~\ref{sec:ClassAssemb} we give the classical and \texttt{OptV1} algorithms.
In Section~\ref{sec:CodeOptV2} we present the vectorized
\texttt{OptV2} and \texttt{\OptVec} algorithms for a generic sparse matrix and $\Pk{1}$ finite elements, with
  the application to the assemblies of
the matrices of Section~\ref{sec:Notations}.
A similar version called \texttt{\OptVecS} for symmetric matrices is also given.
A first step towards finite elements of higher order is
deferred to Appendix~\ref{app:ProofLem}.
In Section~\ref{sec:VecCase} we consider the extension to the vector case with an application to linear elasticity.
In Section~\ref{sec:NumRes}, benchmark results illustrate the performance of the algorithms in the Matlab,
Octave and Python languages. First, we show a comparison between the classical,
\texttt{OptV1}, \texttt{OptV2}, \texttt{\OptVec} and \texttt{\OptVecS} versions. Then we compare the performances of the \texttt{\OptVecS} version
to those obtained with a compiled language (using SuiteSparse~\cite{Davis:SSP:2012} in C language),
the latter being well-known to run at high speed and serving as a reference.
A comparison is also given with FreeFEM++~\cite{Hecht:NDFF:2012} as a simple and reliable finite element software.
We also show in Matlab and Octave a comparison of the \texttt{\OptVecS}  algorithm and
the codes given in~\cite{Chen:PFE:2011,Chen:iFEM:2013,Hannukainen:IFE:2012,Rahman:FMA:2011}.

All the computations are done on our reference
computer\footnote{2 x Intel Xeon E5-2630v2 (6 cores) at 2.60Ghz, 64Go RAM}
with the releases R2014b for Matlab, 3.8.1 for Octave, 3.4.0 for Python and 3.31 for FreeFEM++.
The Matlab/Octave and Python codes may be found
in~\cite{CJS:Software:2015}.

\section{Statement of the problem and notation}\label{sec:ProblemAndNotations}
\label{sec:Notations}

In this article we consider the assembly of the standard sparse matrices
(e.g. weighted mass, stiffness and elastic stiffness matrices)
  arising from the $\Pk{1}$ finite element discretization of partial differential equations
  (see e.g.~\cite{Ciarlet:TFE:2002,Quarteroni:NAP:2008})
in a bounded domain $\DOM$ of $\R^d$ ($d\geq 1$).

We suppose that $\DOM$ is equipped with a mesh $\Th$
(locally conforming) as described in Table~\ref{table:notations}.
We suppose that the elements belonging to the mesh are $d$-\simplices.
We introduce the finite dimensional space
 $\XPk{1}=
\{v \in {\cal C}^0(\overline \DOMH),
\ \  {v}_{|\simplex} \in \Pk{1}(\simplex), \ \forall \simplex \in \Th \}$
where $\DOMH=\bigcup_{\simplex \in \Th} \simplex$ and
$\Pk{1}(\simplex)$ denotes the space of all polynomials over $\simplex$
and of total degree less than or equal to~$1$.
Let $\q^j$, $j=1,...,\nq$ be a vertex of~$\DOMH$, with $\nq=dim(\XPk{1})$.
The space $\XPk{1}$ is spanned by the $\Pk{1}$ Lagrange basis functions $\{\BasisFunc_i\}_{i\in\ENS{1}{\nq}}$ in
$\mathbb{R}^d$, where $\BasisFunc_i(\q^j)=\delta_{ij}$, with  $\delta_{ij}$ the Kronecker delta.
\\
We consider two examples of finite element matrices: the weighted mass matrix
$\MassW{w}$, with $w\in\LInfD{\DOM}$, defined by
\begin{equation}\label{eq:massw}
\MassW{w}_{i,j}
=\int_\DOMH w\BasisFunc_j\BasisFunc_i d\q, \quad \forall (i,j)\in \{1,...,\nq\}^2,
\end{equation}
and the stiffness matrix $\Stiff$ given by
\begin{equation}\label{eq:stiffmat}
\Stiff_{i,j}
=\int_\DOMH \DOT{\GRAD\BasisFunc_j}{\GRAD\BasisFunc_i}d\q, \quad \forall (i,j)\in \{1,...,\nq\}^2.
\end{equation}
Note that on the $k$-th element $\simplex=T_k$ of $\Th$ we have
\begin{equation}
\forall \il\in \ENS{1}{d+1},\ \ \ {\BasisFunc_{i}}_{|T_k}=\BaryCoor_\il, \ \text{with } i = \me(\alpha, k),
\end{equation}
where $(\BaryCoor_\il)_{\il\in \ENS{1}{d+1}}$ are the barycentric coordinates
  (i.e the local $\Pk{1}$ Lagrange basis functions) of $\simplex$, and $\me$ is the connectivity array
(see Table~\ref{table:notations}).
The matrices $\MassW{w}$ and $\Stiff$ can be assembled efficiently
with a vectorized algorithm proposed in Section~\ref{sec:CodeOptV2}, which uses
the following formula (see e.g.~\cite{Quarteroni:NMDP:2014})
\begin{equation}\label{MagicFormula}
\int_\simplex \prod_{i=1}^{d+1} \BaryCoor_i^{n_i} d\q = \displaystyle
d!|\simplex|\frac{\displaystyle\prod_{i=1}^{d+1} n_i! }{\displaystyle(d+\sum_{i=1}^{d+1}n_i )!}
\end{equation}
where $|\simplex|$ is the volume of $\simplex$ and $n_i \in \N$.

\begin{remark}
The (non-vectorized or vectorized) finite element assembly algorithms presented in this article
may be adapted to compute matrices associated to the bilinear form~\eqref{eq:EDP}.
\end{remark}
\begin{remark}
These algorithms apply to finite element methods of higher order. Indeed, one can express
the $\Pk{k}$-Lagrange basis functions ($k \ge 2$) as polynomials in $\BaryCoor_i$ variable
  and then use formula~\eqref{MagicFormula}.
In Appendix~\ref{sec:PkFEM} we give a first step to obtain a vectorized algorithm for
$\Pk{k}$ finite elements.
\end{remark}

In the remainder of this article, we will use the following notations to describe the triangulation $\Th$
of $\DOM$:
\begin{table}[H]
\begin{center}
\begin{tabular}{lccl}
\hline
\textbf{name} & \textbf{type} & \textbf{dimension} & \textbf{description}\\
\hline
$d$ & integer & 1 & dimension of \simplices of $\Th$\\
$\nq$ & integer & 1 & number of vertices of $\Th$\\
$\nme$ & integer & 1 & number of mesh elements in $\Th$\\
$\q$   & double  &$d\times \nq$ &
\begin{minipage}[t]{5.9cm}
array of vertex coordinates
\end{minipage}\\
$\me$  & integer & $(d+1)\times \nme$ &
\begin{minipage}[t]{5.9cm}
connectivity array
\end{minipage}\\
$\areas$  & double & $1\times \nme$ &
\begin{minipage}[t]{5.9cm}
array of simplex volumes
\end{minipage}\\
\hline
\end{tabular}
\captionof{table}{Data structure associated to the mesh $\Th$}\label{table:notations}
\end{center}
\end{table}
In Table~\ref{table:notations}, for $\al\in\{1,\hdots,d\}$,
${\q}(\al,j)$ represents the $\al$-th coordinate of the $j$-th vertex, $j\in\{1,\hdots,\nq\}.$
The $j$-th vertex will be also denoted by $\q^j$. The term $\me(\jl,k)$ is the storage index of the $\jl$-th vertex
of the $k$-th element, in the array~$q$, for $\jl\in\{1,...,d+1\}$ and $k\in\{1,\hdots,\nme\}$.

We also provide below some common functions and operators of the vectorized algorithmic language used in this article which generalize the operations on scalars to higher dimensional arrays, matrices and vectors:
\vspace{1mm}

{\small
\begin{tabular}{ll}
 $\MAT{A}\gets \MAT{B}$ & Assignment\\
 $\MAT{A}*\MAT{B}$      & matrix multiplication,\\
 $\MAT{A} \MultVec \MAT{B}$     & element-wise multiplication,\\
 $\MAT{A} \DivVec \MAT{B}$     & element-wise division,\\
 $\MAT{A}(:)$     &  all the elements of $\MAT{A}$, regarded as a single column.\\
 $[,]$      & Horizontal concatenation,\\
 $[;]$      & Vertical concatenation,\\
 $\MAT{A}(:,J)$   & $J$-th column of $\MAT{A}$,\\
 $\MAT{A}(I,:)$   & $I$-th row of $\MAT{A}$,\\
 $\FNameStd{Sum}(\MAT{A},dim)$ & sums along the dimension $dim$,\\
 $\mathbb{I}_n$ & $n$-by-$n$ identity matrix,\\
  $\mathds{1}_{m\times n}$ (or $\mathds{1}_{n}$)& $m$-by-$n$ (or $n$-by-$n$) matrix or sparse matrix of ones,\\
  $\mathds{O}_{m\times n}$ (or $\mathds{O}_{n}$) & $m$-by-$n$ (or $n$-by-$n$) matrix or sparse matrix of zeros,\\
  $\FNameStd{ones}(n_1,n_2,...,n_\ell)$ & $\ell$ dimensional array of ones,\\
  $\FNameStd{zeros}(n_1,n_2,...,n_\ell)$ & $\ell$ dimensional array of zeros.
\end{tabular}
}

\section{Standard finite element assemblies}
\label{sec:ClassAssemb}
In this section we consider the $\Pk{1}$ finite element assembly of a generic $\nq$-by-$\nq$ sparse matrix $\MAT{M}$
with its corresponding $(d+1)$-by-$(d+1)$ local matrix $\MAT{E}$ (also denoted by
$\MAT{E}(\simplex)$ when referring to an element $\simplex \in \Th$).
For $\simplex=T_k$, the $(\il,\jl)$-th entry of $\MAT{E}(T_k)$ is denoted by $e^k_{\il,\jl}$.

In Algorithm~\ref{algo:ClassicalAssembly},
we recall the classical finite element assembly method for calculating $\MAT{M}$.
In this algorithm, an $\nq$-by-$\nq$ sparse matrix $\MAT{M}$ is first declared, then
the contribution of each element $T_k \in \Th$, given by a function \texttt{ElemMat}, is added to the matrix $\MAT{M}$.
These successive operations are very expensive due to a suboptimal
use of the \texttt{sparse} function.

A first optimized, non-vectorized,
version (called \texttt{OptV1}), suggested in \cite{Davis:DMS:2006},
is based on the use of the \texttt{sparse} function:\\[2mm]
\centerline{\texttt{M $\leftarrow$ sparse(Ig,Jg,Kg,m,n);}}\\[2mm]
This command returns an
\texttt{m} $\times$\texttt{n}
sparse matrix $M$
such that\\[2mm]
\centerline{\texttt{M(Ig(k),Jg(k))} $\leftarrow$ \texttt{M(Ig(k),Jg(k)) + Kg(k)}.}\\[2mm]
The vectors \texttt{Ig}, \texttt{Jg} and \texttt{Kg} have the same
length.  The zero elements of \texttt{K} are not taken into account
and the elements of \texttt{Kg} having the same indices in
\texttt{Ig} and \texttt{Jg} are summed.\\
Examples of languages containing a sparse function are given below
\begin{itemize}
  \item[$\bullet$] Python (\textit{scipy.sparse} module) : \\
    \centerline{\texttt{M=sparse.<format>\_matrix((Kg,(Ig,Jg)),shape=(m,n))}}
    where \texttt{<format>} is the sparse matrix format (e.g. \texttt{csc}, \texttt{csr}, \texttt{lil}, ...),
  \item[$\bullet$] Matlab : \texttt{M=sparse(Ig,Jg,Kg,m,n)}, only \texttt{csc} format,
  \item[$\bullet$] Octave : \texttt{M=sparse(Ig,Jg,Kg,m,n)}, only \texttt{csc} format,
  \item[$\bullet$] Scilab : \texttt{M=sparse([Ig,Jg],Kg,[m,n])}, only \texttt{row-by-row} format.
  \item[$\bullet$] C  with \textit{SuiteSparse}~\cite{Davis:SSP:2012}
    \item[$\bullet$] CUDA  with \textit{Thrust}~\cite{Nvidia:Thrust} and \textit{Cusp}~\cite{Nvidia:Cusp} libraries
 \end{itemize}

\begin{figure}[H]
  \imageps{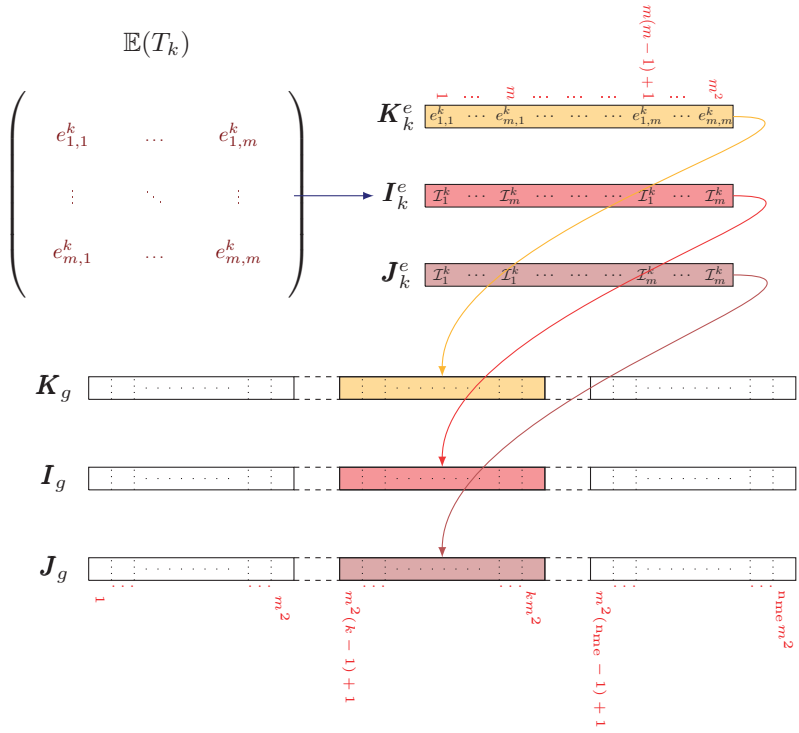}{1}
  \caption{Insertion of a local matrix into global 1d-arrays - \texttt{OptV1} version,
    where $m=d+1,$ $\mathcal{I}^k_l=\me(l,k)$.}
  \label{fig:optV1assembly}
\end{figure}

The \texttt{OptV1} version
consists in computing and storing all elementary contributions first and then using them to
generate the sparse matrix $\MAT{M}$.
The main idea is to create three global 1d-arrays $\vecb{K}_g$, $\vecb{I}_g$ and $\vecb{J}_g$ of length $(d+1)^2\nme$,
which store the local matrices as well as the
position of their elements in the global matrix as shown on Figure~\ref{fig:optV1assembly}.
To create the arrays $\vecb{K}_g$, $\vecb{I}_g$ and $\vecb{J}_g$,
we define three local arrays $\vecb{K}^e_k,$ $\vecb{I}^e_k$ and
$\vecb{J}^e_k$ of length $(d+1)^2$ obtained from the $(d+1)$-by-$(d+1)$ local matrix $\MAT{E}(T_k)$
as follows:
\begin{center}
\begin{tabular}{lcl}
$\vecb{K}^e_k$ &:& elements of the matrix $\MAT{E}(T_k)$ stored column-wise,\\
$\vecb{I}^e_k$ &:& global row indices associated to the elements stored in $\vecb{K}^e_k$,\\
$\vecb{J}^e_k$ &:& global column indices associated to the elements stored in $\vecb{K}^e_k.$
\end{tabular}
\end{center}
Using $\vecb{K}^e_k$, $\vecb{I}^e_k$, $\vecb{J}^e_k$ and a loop over the mesh elements $T_k,$ one may calculate
the global arrays $\vecb{I}_g$, $\vecb{J}_g$ and $\vecb{K}_g$.
The corresponding \texttt{OptV1} algorithm is given in Algorithm~\ref{Assemblage_OptV1_algo}.\\

Numerical experiments in Section~\ref{sub:CompVersions} and in Tables~\ref{tab:Stiff3D_Merge} and~\ref{tab:StiffElas2D_Merge} show that the \texttt{OptV1} algorithm is more efficient than the classical one.
 The inefficiency of the classical (\texttt{base}) version compared to the \texttt{OptV1} version
 is mainly due to the repetition of element insertions into the sparse structure and to some
dynamic reallocation troubles that may also occur.

However, the \texttt{OptV1} algorithm still uses a loop over the elements.
To improve the efficiency of this algorithm, we propose in the next section other optimized versions,
in a vectorized form: the main loop over the elements, which increases with the size of the mesh,
is vectorized. The other loops (which are independent of the mesh size and with few iterations)
will not necessarily be vectorized.

\begin{center}
\begin{minipage}[t]{0.49\textwidth}
\begin{algorithm}[H]
\captionsetup{font=footnotesize}
\caption{(\texttt{base}) - Classical assembly  }
\label{algo:ClassicalAssembly}
\begin{algorithmic}[1]
\State $\MAT{M} \gets \mathds{O}_{\nq}$ \Comment{{\tiny Sparse matrix}}
\For{k}{1}{\nme}
  \State $\MAT{E}\gets$ \texttt{ElemMat}($\areas(k),\hdots$)
  \For{\il}{1}{d+1}
    \State $i\gets \me(\il,k)$
    \For{\jl}{1}{d+1}
      \State $j\gets \me(\jl,k)$
      \State $ \MAT{M}_{i,j} \gets \MAT{M}_{i,j} +\MAT{E}_{\il,\jl}$
    \EndFor
  \EndFor
\EndFor
\end{algorithmic}
\end{algorithm}
\end{minipage}\hspace{0.005\textwidth}
\begin{minipage}[t]{0.49\textwidth}
\begin{algorithm}[H]
\captionsetup{font=footnotesize}
\caption{(\texttt{OptV1}) - Optimized and non-vectorized assembly}
\label{Assemblage_OptV1_algo}
 \begin{algorithmic}[1]
   \State $\vecb{K}_g \gets \vecb{I}_g \gets \vecb{J}_g \gets \FNameStd{zeros}((d+1)^2 \nme,1)$
  \State $l\gets 1$
 \For{k}{1}{\nme}
   \State $\MAT{E}\gets$ \texttt{ElemMat}($\areas(k),\hdots$)
   \For{\jl}{1}{d+1}
     \For{\il}{1}{d+1}
	\State $\vecb{I}_g(l)\gets \me(\il,k)$
	\State $\vecb{J}_g(l)\gets \me(\jl,k)$
	\State $\vecb{K}_g(l)\gets \MAT{E}(\il,\jl)$
	\State $l \gets l+1$
     \EndFor
   \EndFor
 \EndFor
 \State $\MAT{M} \gets $\texttt{sparse}($\vecb{I}_g$,$\vecb{J}_g$,$\vecb{K}_g$,$\nq$,$\nq$)
 \end{algorithmic}
\end{algorithm}
\end{minipage}
\end{center}

\section{Optimized finite element assembly}
\label{sec:CodeOptV2}
In this section we present optimized algorithms, only available in vector languages.
In the first algorithm, \texttt{OptV2},
the idea is to vectorize the main loop over the elements by defining
the two-dimensional arrays $\MAT{K}_g,$ $\MAT{I}_g$ and $\MAT{J}_g$ of size $(d+1)^2$-by-$\nme$
  which store all the local matrices as well as their positions in the global matrix.
Then, as for the \texttt{OptV1} version, the matrix assembly is obtained with the sparse function:
\vspace{-0.5mm}
\begin{center}
  \texttt{M $\leftarrow$ sparse($\MAT{I}_g(:)$,$\MAT{J}_g(:)$,$\MAT{K}_g(:)$,$\nq$,$\nq$);}
\end{center}

\begin{figure}[H]
  \imageps{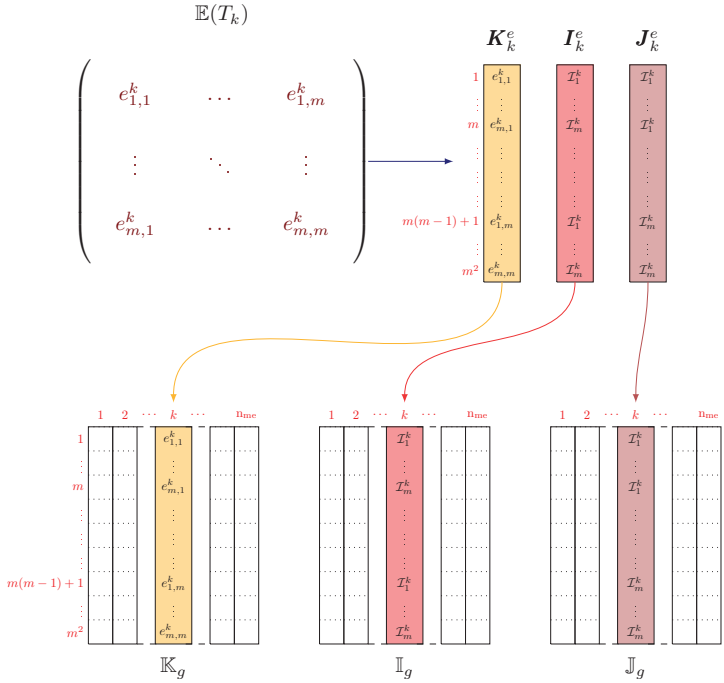}{0.8}
  \caption{Insertion of a local matrix into global 2D-arrays}
  \label{fig:optV2assemblyintro}
\end{figure}

A non-vectorized approach inspired by \texttt{OptV1} is as follows: for each mesh element $T_k$,
the $k$-th column of the global arrays $\MAT{K}_g,$ $\MAT{I}_g$ and $\MAT{J}_g$
is filled with the local arrays $\vecb{K}^e_k,$ $\vecb{I}^e_k$, $\vecb{J}^e_k$ respectively,
  as shown in Figure~\ref{fig:optV2assemblyintro}.
Thus, $\MAT{K}_g,$ $\MAT{I}_g$ and $\MAT{J}_g$ are defined by:
$\forall k \in \ENS{1}{\nme},$ $\forall l \in\ENS{1}{(d+1)^2}$,
\begin{eqnarray*}
\MAT{K}_g(l,k)=\vecb{K}^e_k(l),\quad\quad
\MAT{I}_g(l,k)=\vecb{I}^e_k(l)
,\quad\quad
\MAT{J}_g(l,k)=\vecb{J}^e_k(l).
\end{eqnarray*}
A natural way to calculate these three arrays is column-wise. In that case, for each array one needs
to compute $\nme$ columns.

\begin{figure}[H]
  \imageps{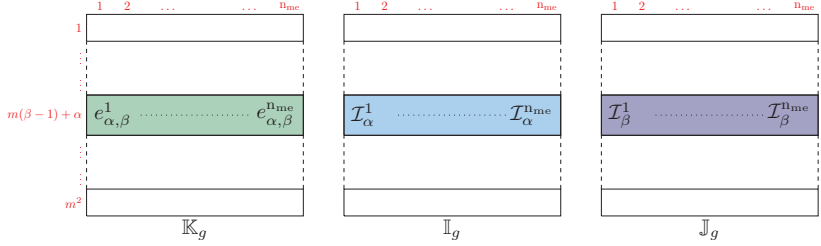}{0.9}
  \caption{Row-wise operations on global 2D-arrays}
  \label{fig:optV2assembly}
\end{figure}

The \texttt{OptV2} method consists in calculating these arrays row-wise.
In that case, for each array one needs to calculate $(d+1)^2$ rows
(where $d$ is independent of the number of mesh elements).
This vectorization method is represented in Figure~\ref{fig:optV2assembly}.
We first suppose that for $\il$ and $\jl$ fixed, we can vectorize
the computation of~$e^k_{\il,\jl},$ for all $k\in\ENS{1}{\nme}.$ This vectorization procedure,
denoted by $\FName{vecElem}(\il,\jl,\hdots)$, returns a 1d-array containing these $\nme$ values.
We will describe it in detail for some examples in Sections~\ref{subsec:WeightedMass} and~\ref{subsec:Stiff}.
Then we obtain the following algorithm
%
\begin{algorithm}[H]
\captionsetup{font=footnotesize}
\caption{(\texttt{OptV2}) - Optimized and vectorized assembly}
\label{algo:AssemblyGenP1}
\ifdefined\IsInputOutput
\begin{minipage}{14.cm}
\footnotesize{\textbf{Input :}\\
  \begin{tabular}[t]{lcl}
     $\me$&:& $(d+1)$-by-$\nme$ connectivity array\\
     $\nq$&:& number of vertices\\
  \end{tabular}\\
\textbf{Output :}\\
  \begin{tabular}{lcl}
    $\MAT{M}$ & : & $\nq$-by-$\nq$ sparse matrix
  \end{tabular}\\
}
\end{minipage}
\fi
\begin{algorithmic}[1]
\Function{$\MAT{M} \gets $\FNameDef{AssemblyGenP1OptV2}}{$\me,\nq,\hdots$}
\State $\Kg \gets \Ig \gets \Jg \gets \FNameStd{zeros}((d+1)^2,\nme)$\Comment{$(d+1)^2$-by-$\nme$ 2d-arrays}
\State $l\gets 1$
\For{\jl}{1}{d+1}
  \For{\il}{1}{d+1}
   \State $ \Kg(l,:) \gets \FName{vecElem}(\il,\jl,\hdots)$ \label{algo:Kg:line}
   \State $ \Ig(l,:) \gets  \me(\il,:)$
   \State $ \Jg(l,:) \gets  \me(\jl,:)$
   \State $ l \gets l+1$
  \EndFor
\EndFor
\State $\MAT{M} \gets \FNameStd{Sparse}(\Ig(:),\Jg(:),\Kg(:),\nq,\nq)$
\EndFunction
\end{algorithmic}
\end{algorithm}

Algorithm~\ref{algo:AssemblyGenP1} is efficient in terms of computation time (see Section~\ref{sub:CompVersions}).
However it is memory consuming due to the size of the arrays $\MAT{I}_g$, $\MAT{J}_g$ and $\MAT{K}_g$.
Thus a variant (see~\cite{Chen:iFEM:2013,Koko:SP:2007} for dimension 2 or 3 in Matlab) consists in using the sparse command inside the loops
(i.e. for each component of all element matrices). This method, called~\texttt{\OptVec}, is
given in Algorithm~\ref{algo:AssemblyGenP1OptVec}.
%
\begin{algorithm}[H]
\captionsetup{font=footnotesize}
\caption{(\texttt{\OptVec}) - Optimized and vectorized assembly (less memory consuming)}
\label{algo:AssemblyGenP1OptVec}
\ifdefined\IsInputOutput
\begin{minipage}{14.cm}
\footnotesize{\textbf{Input :}\\
  \begin{tabular}[t]{lcl}
     $\me$&:& $(d+1)$-by-$\nme$ connectivity array\\
     $\nq$&:& number of vertices\\
  \end{tabular}\\
\textbf{Output :}\\
  \begin{tabular}{lcl}
    $\MAT{M}$ & : & $\nq$-by-$\nq$ sparse matrix
  \end{tabular}\\
}
\end{minipage}
\fi
\begin{algorithmic}[1]
\Function{$\MAT{M} \gets $\FNameDef{AssemblyGenP1\OptVec}}{$\me,\nq,\hdots$}
\State $\MAT{M} \gets \MAT{O}_{\nq}$ \Comment{$\nq$-by-$\nq$ sparse matrix}
\For{\jl}{1}{d+1}
  \For{\il}{1}{d+1}
   \State $ \vecb{K}_g \gets \FName{vecElem}(\il,\jl,\hdots)$ \label{algo:Kg:line}
   \State $\MAT{M} \gets \MAT{M} + \FNameStd{Sparse}(\me(\il,:),\me(\jl,:),\vecb{K}_g,\nq,\nq)$
  \EndFor
\EndFor
\EndFunction
\end{algorithmic}
\end{algorithm}

\vspace{-1mm}

For a symmetric matrix, the performance can be improved by using a symmetrized version  of
  \texttt{\OptVec} (called \texttt{\OptVecS}), given in Algorithm~\ref{algo:AssemblyGenP1OptVecS}.
More precisely,
in the lines \ref{algo:AssemblyGenP1OptVecS:l01}-\ref{algo:AssemblyGenP1OptVecS:l02}
of this algorithm, we build a non-triangular sparse matrix which
contains the contributions of the strictly upper parts of all the element matrices.
In line \ref{algo:AssemblyGenP1OptVecS:l03}
the strictly lower part contributions are added using the symmetry of the element matrices. Then in lines
\ref{algo:AssemblyGenP1OptVecS:l04}-\ref{algo:AssemblyGenP1OptVecS:l05} the contributions of
the diagonal parts of the element matrices are added.

\vspace{-1mm}

\begin{algorithm}[H]
\captionsetup{font=footnotesize}
\caption{(\texttt{\OptVecS}) -  Symmetrized version of \texttt{\OptVec}}
\label{algo:AssemblyGenP1OptVecS}
\ifdefined\IsInputOutput
\begin{minipage}{14.cm}
\footnotesize{\textbf{Input :}\\
  \begin{tabular}[t]{lcl}
     $\me$&:& $(d+1)$-by-$\nme$ connectivity array\\
     $\nq$&:& number of vertices\\
  \end{tabular}\\
\textbf{Output :}\\
  \begin{tabular}{lcl}
    $\MAT{M}$ & : & $\nq$-by-$\nq$ sparse matrix
  \end{tabular}\\
}
\end{minipage}
\fi
\begin{algorithmic}[1]
\Function{$\MAT{M} \gets $\FNameDef{AssemblyGenP1\OptVecS}}{$\me,\nq,\hdots$}
\State $\MAT{M} \gets \MAT{O}_{\nq}$ \Comment{$\nq$-by-$\nq$ sparse matrix}
\For{\il}{1}{d+1}\label{algo:AssemblyGenP1OptVecS:l01}
  \For{\jl}{\il+1}{d+1}
   \State $ \vecb{K}_g \gets \FName{vecElem}(\il,\jl,\hdots)$  
   \State $\MAT{M} \gets \MAT{M} + \FNameStd{Sparse}(\me(\il,:),\me(\jl,:),\vecb{K}_g,\nq,\nq)$
  \EndFor
\EndFor\label{algo:AssemblyGenP1OptVecS:l02}
\State $\MAT{M} \gets \MAT{M} + \MAT{M}^t$ \label{algo:AssemblyGenP1OptVecS:l03}
\For{\il}{1}{d+1}\label{algo:AssemblyGenP1OptVecS:l04}
  \State $ \vecb{K}_g \gets \FName{vecElem}(\il,\il,\hdots)$ 
  \State $\MAT{M} \gets \MAT{M} + \FNameStd{Sparse}(\me(\il,:),\me(\il,:),\vecb{K}_g,\nq,\nq)$
\EndFor\label{algo:AssemblyGenP1OptVecS:l05}
\EndFunction
\end{algorithmic}
\end{algorithm}

\vspace{-1mm}

In the following, our objective is to show using examples how to vectorize
the computation of $\vecb{K}_g$ (i.e. how to obtain the $\FName{vecElem}$ function
in algorithms \texttt{OptV2}, \texttt{\OptVec} and \texttt{\OptVecS}).
More precisely for the examples derived from~\eqref{eq:EDP},
the calculation of $\vecb{K}_g$
only depends on the local basis functions and/or their gradients and one may need to calculate them
on all mesh elements.
For $\Pk{1}$ finite elements, these gradients are constant on each $d$-simplex $\simplex=T_k.$
Let $\MAT{G}$ be the 3D array of size $\nme$-by-$(d+1)$-by-$d$ defined by
\begin{eqnarray}\label{eq:defGalpha}
\MAT{G}(k,\il,:)=\GRAD\BasisFunc_{\il}^k(\q),\ \ \forall \il\in\ENS{1}{d+1},\ \forall k\in\ENS{1}{\nme}.
\end{eqnarray}
In Appendix~\ref{Appendix:Gradients}, we give a
vectorized function called \FName{GradientVec} (see Algorithm~\ref{algo:GradientVec})
which computes $\MAT{G}$ in arbitrary dimension.
Once the gradients are computed, the local matrices are calculated using
the formula (\ref{MagicFormula}). For simplicity, in the following we consider the \texttt{\OptVec} version.
The vectorization of the computation of $\vecb{K}_g$ is shown using the two examples introduced in Section~\ref{sec:Notations}.

\subsection{Weighted mass matrix assembly}\label{subsec:WeightedMass}
The local weighted mass matrix $\MassWElem{w}(\simplex)$ is given by
\begin{equation}\label{eq:masswelemmat}
\MassWElem{w}_{\il,\jl}(\simplex)
=\int_{\simplex} w\BaryCoor_\jl\BaryCoor_\il d\q, \quad \forall (\il,\jl) \in \{1,...,d+1\}^2,
\end{equation}
with $w\in\LInfD{\DOM}$.
Generally, this matrix cannot be computed exactly and one has to use a quadrature formula.
  In the following, we choose to approximate $w$ by $w_h=\PihK{1}(w)$ where
$\PihK{1}(w) =\sum_{\gamma=1}^{d+1}  w(\q^{\gamma}) \BaCo_{\gamma}$ is the $\Pk{1}$ Lagrange interpolation of $w$.
Then using~\eqref{MagicFormula}, we have the quadrature formula for \eqref{eq:masswelemmat}
\begin{equation}\label{eq:MatElemMassWP1}
\int_\simplex  \PihK{1}(w) \BaCo_{\alpha} \BaCo_{\beta}d\q
=\frac{d!}{(d+3)!}|K|(1+\delta_{\il,\jl})
(w^s+w(\q^{\il})+w(\q^{\jl})),
\end{equation}
where $w^s=\sum_{\gamma=1}^{d+1}  w(\q^{\gamma})$.
Using~\eqref{eq:MatElemMassWP1} we vectorize the assembly of the approximate
weighted mass matrix~\eqref{eq:massw} as
shown in Algorithm~\ref{algo:AssemblyMassWP1}.
\begin{algorithm}[H]
\captionsetup{font=footnotesize}
\caption{(\texttt{\OptVec}) - Weighted mass matrix assembly}
\label{algo:AssemblyMassWP1}
\ifdefined\IsInputOutput
\begin{minipage}{14.cm}
\footnotesize{\textbf{Input :}\\
  \begin{tabular}[t]{lcl}
     $\me$&:& $(d+1)$-by-$\nme$ connectivity array\\
     $\q$ & :& $\dd$-by-$\nq$ array of vertices coordinates\\
     $w$ &: & weight function
  \end{tabular}\\
\textbf{Output :}\\
  \begin{tabular}{lcl}
    $\MAT{M}$ & : & $\nq$-by-$\nq$ sparse matrix
  \end{tabular}\\
}
\end{minipage}
\fi
\begin{algorithmic}[1]
\Function{$\MAT{M} \gets $\FNameDef{AssemblyMassWP1\OptVec}}{$\me,\q,\volumes,w$}
\State $\vecb{w} \gets w(\q)$ \Comment{1d-array of size $\nq$}
\State $\MAT{W} \gets \vecb{w}(\me)$ \Comment{ $(d+1)$-by-$\nme$ 2d-array}
\State $\vecb{w}^s \gets \FNameStd{Sum}(\MAT{W},1)$ \Comment{1d-array of size $\nme$}\label{algo:AssemblyMassWP1:ws}
\State $\MAT{M} \gets \MAT{O}_{\nq}$ \Comment{$\nq$-by-$\nq$ sparse matrix}
\For{\il}{1}{d+1}
  \For{\jl}{1}{d+1}
   \State $ \vecb{K}_g \gets \frac{d!}{(d+3)!}(1+\delta_{\il,\jl}) * \volumes \MultVec 
            (\vecb{w}^s \PlusVec \MAT{W}(\il,:) \PlusVec \MAT{W}(\jl,:))$  \label{algo:AssemblyMassWP1:Kg}
   \State $\MAT{M} \gets \MAT{M} + \FNameStd{Sparse}(\me(\il,:),\me(\jl,:),\vecb{K}_g,\nq,\nq)$
  \EndFor
\EndFor
\EndFunction
\end{algorithmic}
\end{algorithm}

\noindent
Line~\ref{algo:AssemblyMassWP1:Kg} of Algorithm~\ref{algo:AssemblyMassWP1} corresponds to
the vectorization of formula~\eqref{eq:MatElemMassWP1} and is carried out as
follows: first we set $\vecb{w}\in\R^\nq$ such that $\vecb{w}(i)=w(\q^i),$
$1\le i \le \nq$, or in a vectorized form $ \vecb{w} \gets w(\q)$. Then we compute
the array $\MAT{W}$ of size $(d+1)$-by-$\nme$ containing, for each $d$-simplex, the values of $w$ at its vertices:
$\MAT{W}(\il,k)=w(\q^{\me(\il,k)})$ or in vectorized form $\MAT{W} \gets \vecb{w}(\me).$ We now calculate
$\vecb{w}^s\in\R^\nme$ which contains, for each $d$-simplex, the sum of the values of $w$
at its vertices, i.e. we sum $\MAT{W}$ over the rows and obtain line~\ref{algo:AssemblyMassWP1:ws} of
Algorithm~\ref{algo:AssemblyMassWP1}.
Then, formula~\eqref{eq:MatElemMassWP1} may be vectorized to obtain
line~\ref{algo:AssemblyMassWP1:Kg} in Algorithm~\ref{algo:AssemblyMassWP1}.
\begin{remark}
Note that formula~\eqref{eq:MatElemMassWP1} is exact if $w$ is a polynomial of degree~$1$ on~$\simplex$.
Moreover, if $w$ is constant, we get the mass matrix (up to the constant~$w$).
Other quadrature rules could be used to approximate the integral
in~\eqref{eq:masswelemmat}
without changing the principle of Algorithm~\ref{algo:AssemblyMassWP1}.
\end{remark}

\begin{remark}
Algorithm~\ref{algo:AssemblyMassWP1} can be applied to meshes composed of $n$-\simplices (for $n\leq d$) and
may be used to compute Neumann or Robin boundary terms.
\end{remark}

\subsection{Stiffness matrix assembly}\label{subsec:Stiff}
The local stiffness matrix $\StiffElem(\simplex)$
  is given, for all $(\il,\jl) \in \{1,...,d+1\}^2$, by
\begin{equation}\label{eq:stiffelemmat}
\StiffElem_{\il,\jl}(\simplex)
=\int_{\simplex} \DOT{\GRAD\BaryCoor_\jl}{\GRAD\BaryCoor_\il}d\q
=|K|\DOT{\GRAD\BaryCoor_\jl}{\GRAD\BaryCoor_\il}.
\end{equation}
To obtain the right-hand side of \eqref{eq:stiffelemmat} we use the fact that
the gradients of the local basis functions
are constant on each $d$-simplex. The gradients are computed with the vectorized function \FName{GradientVec} of Algorithm~\ref{algo:GradientVec}.
Then the vectorized assembly Algorithm~\ref{algo:AssemblyStiffP1} easily follows.

\begin{algorithm}[H]
\captionsetup{font=footnotesize}
\caption{(\texttt{\OptVec}) - Stiffness matrix assembly}
\label{algo:AssemblyStiffP1}
\ifdefined\IsInputOutput
\begin{minipage}{14.cm}
\footnotesize{\textbf{Input :}\\
  \begin{tabular}[t]{lcl}
     $\me$&:& $(d+1)$-by-$\nme$ connectivity array\\
     $\q$ & :& $\dd$-by-$\nq$ array of vertices coordinates\\
     $\volumes$ &: & $1$-by-$\nme$ array of volumes 
  \end{tabular}\\
\textbf{Output :}\\
  \begin{tabular}{lcl}
    $\MAT{S}$ & : & $\nq$-by-$\nq$ sparse matrix
  \end{tabular}\\
}
\end{minipage}
\fi
\begin{algorithmic}[1]
\Function{$\MAT{M} \gets $\FNameDef{AssemblyStiffP1\OptVec}}{$\me,\q,\volumes$}
\State $\MAT{G}\gets \FName{GradientVec}(\q,\me)$ 
\State $\MAT{M} \gets \MAT{O}_{\nq}$ \Comment{$\nq$-by-$\nq$ sparse matrix}
\For{\il}{1}{d+1}
  \For{\jl}{1}{d+1}
    \State $ \vecb{K}_g \gets \FNameStd{zeros}(1,\nme)$ 
    \For{i}{1}{d}
      \State $ \vecb{K}_g \gets \vecb{K}_g \PlusVec \MAT{G}(:,\jl,i) \MultVec \MAT{G}(:,\il,i)$
    \EndFor
    \State $ \vecb{K}_g \gets \vecb{K}_g  \MultVec \volumes$
    \State $\MAT{M} \gets \MAT{M} + \FNameStd{Sparse}(\me(\il,:),\me(\jl,:),\vecb{K}_g,\nq,\nq)$
  \EndFor
\EndFor
\EndFunction
\end{algorithmic}
\end{algorithm}

We will now adapt these methods to the vector case with an application to the assembly of the elastic stiffness matrix
in two and three dimensions.

\section{Extension to the vector case}
\label{sec:VecCase}
In this section we present an extension of Algorithms~\ref{algo:AssemblyGenP1} and~\ref{algo:AssemblyGenP1OptVec} to the vector case, i.e for a system of $m$ ($m>1$) partial differential equations such as in elasticity.
  First, we need to introduce some notation:
the space $(\XPk{1})^m$ (where $\XPk{1}$ is defined in Section~\ref{sec:Notations}),
is of dimension $\ndf=m\,\nq$ and
spanned by the vector basis functions
$\{\BasisFuncTwoD_{l,i}\}_{\substack{1\le i \le \nq\\1\le l \le m }}$,
given by
\begin{equation}
\BasisFuncTwoD_{l,i} = \BasisFunc_{i} \vecb{e}_{l},
\end{equation}
where $\{\vecb{e}_{1},\cdots,\vecb{e}_{m}\}$ is the standard basis of $\R^m.$
The \textit{alternate} numbering is chosen for the basis functions.
We use either $\BasisFuncTwoD_{l,i}$ or $\BasisFuncTwoD_{s}$
with $s=(i-1)m+l$ to denote them.
We will consider the assembly of a generic sparse matrix of dimension $\ndof$-by-$\ndof$ defined by
$$\MAT{H}_{r,s}=\int_\DOMH \mathcal{H}(\BasisFuncTwoD_{s},\BasisFuncTwoD_{r})d\q,$$
where $\mathcal{H}$ is a bilinear differential operator of order one.

As in the scalar case, in order to vectorize the assembly of the matrix, one has to
vectorize the computation of the local matrices. To define the local matrix, we introduce the following
notation:
on the $k$-th element $\simplex=T_k$ of $\Th$ we denote by
$\{\BaryCoorVec_{\l,\il}\}_{\substack{1 \le l \le m\\ 1 \le \il \le d+1}}$ the $\ndfe=m(d+1)$ local basis functions
defined by
\begin{equation}\label{def:BaryCoorVec}
\BaryCoorVec_{\l,\il}=\BaryCoor_\il \vecb{e}_{l}.
\end{equation}
We also use notation $\BaryCoorVec_{i}$ with $i=(\il-1)m+l$ to denote  $\BaryCoorVec_{l,\il}.$
By construction,  we have
$\forall l\in\ENS{1}{m},$ $\forall \il\in\ENS{1}{d+1}$
$$
\BasisFuncTwoD_{l,\me(\il,k)}=\BaryCoorVec_{\l,\il}\ \ \mbox{on}\ \simplex=T_k.
$$
Thus, the local matrix $\MAT{H}^e$ on the $d$-simplex~$K$ is of size $\ndfe$-by-$\ndfe$, and is given by
$$\MAT{H}^e_{i,j}=\int_\simplex \mathcal{H}(\BaryCoorVec_{j},\BaryCoorVec_{i})d\q.$$
Then, a classical non-vectorized algorithm is given in Algorithm~\ref{algo:ClassicalAssemblyVectorCase}.
The function~\FName{ElemH} is used to calculate the matrix $\MAT{H}^e$ for a given $d$-simplex~$\simplex$.
As in the scalar case, the vectorized assembly algorithm is based on the use of a function called \FName{vecHe}
which returns the values corresponding to
the $(i,j)$-th entry (with $(i,j)=(m(\il-1)+l,m(\jl-1)+n)$) of the local matrices
$\MAT{H}^e(\simplex),$ for all $\simplex \in \Th$ and for all $l,\il,n,\jl.$
We suppose that this function can be vectorized. Then we obtain the \texttt{OptV2}
vectorized assembly of the matrix~$\MAT{H}$
given in Algorithm~\ref{algo:AssemblyGenP1vector}.
\begin{center}
\begin{minipage}[t]{0.49\textwidth}
\begin{algorithm}[H]
\captionsetup{font=footnotesize}
\caption{(\texttt{base}) - Classical assembly in vector case ($m>1$)}
\label{algo:ClassicalAssemblyVectorCase}
\begin{algorithmic}[1]

\State $\ndof\gets m*\nq$ 
\State $\MAT{H} \gets \mathds{O}_{\ndof}$ \Comment{{\tiny Sparse matrix}}
\For{k}{1}{\nme}
  \State $\MAT{H}^e\gets$ \texttt{ElemH}($\areas(k),\hdots$)
  \For{l}{1}{m}
    \For{n}{1}{m}
      \For{\il}{1}{d+1}
        \State $r\gets m *(\me(\il,k)-1)+l$
        \State $i \gets m *(\il-1) + l$
        \For{\jl}{1}{d+1}
           \State $s\gets m *(\me(\jl,k)-1)+n$
           \State $j \gets m *(\jl-1) + n$
           \State $ \MAT{H}_{r,s} \gets \MAT{H}_{r,s} +\MAT{H}^e_{i,j}$
         \EndFor
       \EndFor
    \EndFor
  \EndFor
\EndFor 
\end{algorithmic}
\end{algorithm}

\end{minipage}
\scalebox{0.97}{\begin{minipage}[t]{0.51\textwidth}
\begin{algorithm}[H]
\captionsetup{font=footnotesize}
\caption{(\texttt{OptV2}) - Optimized assembly  in vector case ($m>1$)}
\label{algo:AssemblyGenP1vector}
\begin{algorithmic}[1]
\State $\ndfe \gets m * (d+1)$
\State $\Kg \gets \Ig \gets \Jg \gets  \FNameStd{zeros}(\ndfes,\nme)$
\State $p\gets 1$
\For{l}{1}{m}
  \For{n}{1}{m}
    \For{\jl}{1}{d+1}
       \For{\il}{1}{d+1}
	  \State $ \Kg(p,:) \gets \FName{vecHe}(l,\il,n,\jl,\hdots)$
	  \State $ \Ig(p,:) \gets  m*(\me(\il,:)-1)+l$
	  \State $ \Jg(p,:) \gets  m*(\me(\jl,:)-1)+n$
	  \State $ p \gets p+1$
      \EndFor
    \EndFor
  \EndFor
\EndFor
\State $\ndof \gets m*\nq$ 
\State $\MAT{H} \gets \tiny{\FNameStd{sparse}(\Ig(:),\Jg(:),\Kg(:),\ndof,\ndof)}$
\end{algorithmic}
\end{algorithm}

\end{minipage}}
\end{center}

As in Section~\ref{sec:CodeOptV2}, although Algorithm~\ref{algo:AssemblyGenP1vector}
is efficient in terms of computation time, it is memory consuming due to the size of the arrays
$\MAT{I}_g$, $\MAT{J}_g$ and $\MAT{K}_g$. Thus a variant consists in using the sparse command inside the loops,
which leads to the extension of the~\texttt{\OptVec} algorithm to the vector case,
given in Algorithm~\ref{algo:AssemblyGenP1vectorOptV4}.
%
\begin{algorithm}[H]
\captionsetup{font=footnotesize}
\caption{(\texttt{\OptVec}) - Optimized assembly in vector case ($m>1$) }
\label{algo:AssemblyGenP1vectorOptV4}
\ifdefined\IsInputOutput
\begin{minipage}{14.cm}
\footnotesize{\textbf{Input :}\\
  \begin{tabular}[t]{lcl}
     $\me$&:& $\ndfe$-by-$\nme$ connectivity array \\
          &  &with $\ndfe=d+1$\\
  \end{tabular}\\
\textbf{Output :}\\
  \begin{tabular}{lcl}
    $\MAT{M}$ & : & $\ndof$-by-$\ndof$ sparse matrix \\
                & &where $\ndof=m \nq.$
  \end{tabular}\\
}
\end{minipage}
\fi
\begin{algorithmic}[1]
\Function{$\MAT{M} \gets $\FNameDef{AssemblyVecGenP1\OptVec}}{$\me,\nq,\hdots$}
\State $\ndof \gets m*\nq$ 
\State $\MAT{M} \gets \MAT{O}_{\ndof}$ \Comment{$\ndof$-by-$\ndof$ sparse matrix}
\For{l}{1}{m}
  \For{\il}{1}{d+1}
    \State $ \vecb{I}_g \gets  m*(\me(\il,:)-1)+l$
    \For{n}{1}{m}
       \For{\jl}{1}{d+1}
	  \State $ \vecb{K}_g \gets \FName{vecHe}(l,\il,n,\jl,\hdots)$
	  \State $ \vecb{J}_g \gets  m*(\me(\jl,:)-1)+n$
	  \State $\MAT{M} \gets \MAT{M} + \FNameStd{Sparse}(\vecb{I}_g,\vecb{J}_g,\vecb{K}_g,\ndof,\ndof)$
      \EndFor
    \EndFor
  \EndFor
\EndFor
\EndFunction
\end{algorithmic}
\end{algorithm}

For a symmetric matrix, the performance can be improved by using a symmetrized version
  of Algorithm~\ref{algo:AssemblyGenP1vectorOptV4} (as in Section~\ref{sec:CodeOptV2}), given in
Algorithm~\ref{algo:AssemblyGenP1vectorOptV4s}.

In the following the vectorized function $\FName{vecHe}$ is detailed for the elastic stiffness matrix in 2D and 3D.

\subsection{Elastic stiffness matrix assembly}
Here we consider sufficiently regular vector fields $\VEC{u}=(u_1,\hdots,u_d):\Omega \rightarrow \R^d$,
  with the associated discrete space $(\XPk{1})^d$, $d=2$ or $3$ (i.e. $m=d$ in that case).

  We consider the elastic stiffness matrix
  arising in linear elasticity when Hooke's law is used and the material is isotropic, under
small strain hypothesis (see for example \cite{Dhatt:FEM:2012}).
This sparse matrix $\StiffElas$ is defined by
\begin{equation}\label{eq:StiffElasMat}
\StiffElas_{l,n}=\int_{\DOMH}\Odv^t(\BasisFuncTwoD_n) \mathbb{C}\Odv(\BasisFuncTwoD_l)d\q, \ \ \forall (l,n)\in\ENS{1}{\ndf}^2,
\end{equation}
where $\Odv$ is the linearized strain tensor given by
   $$
   \Odv(\VEC{u}) = \frac 12 \left(\vecb{\GRAD}(\VEC{u})+\vecb{\GRAD}^t(\VEC{u})\right),
   $$
with
$\Odv= (\epsilon_{11}, \epsilon_{22}, 2 \epsilon_{12} )^t$ in 2D and
$\Odv= (\epsilon_{11}, \epsilon_{22},\epsilon_{33}, 2 \epsilon_{12}, 2 \epsilon_{23},2 \epsilon_{13} )^t$ in 3D,
  with $\epsilon_{ij}(\VEC{u})=\frac 12 \left( \DP{u_i}{x_j}+\DP{u_j}{x_i}\right)$.
The elasticity tensor $\mathbb{C}$ depends on
 the Lamé parameters $\lambda$ and $\mu$ satisfying $\lambda+\mu >0,$ and possibly variable in $\DOM$.
For $d=2$ or $d=3$, the matrix $\mathbb{C}$ is given by
\begin{equation*}
\mathbb{C}=
\begin{pmatrix}
  \lambda \mathds{1}_2 +2\mu \mathbb{I}_2&  \  \mathds{O}_{2\times 1}\\
 \mathds{O}_{1\times 2} &  \mu
\end{pmatrix}_{3\times 3},
\qquad
\mathbb{C}=
\begin{pmatrix}
  \lambda \mathds{1}_3 +2\mu \mathbb{I}_3&   \ \mathds{O}_{3\times 3}\\
 \mathds{O}_{3\times 3} &  \mu \mathbb{I}_3
\end{pmatrix}_{6\times 6}.
\end{equation*}
Formula~(\ref{eq:StiffElasMat}) is related to the Hooke's law
\begin{equation*}
\Ocv=\mathbb{C}\Odv,
\end{equation*}
where $\Ocv$ is the elastic stress tensor.

The vectorization of the assembly of the elastic stiffness matrix~\eqref{eq:StiffElasMat} will be carried out
as in Section~\ref{sec:CodeOptV2}, through the vectorization of the
local elastic stiffness matrix $\StiffElasElem$ given for all $(i,j)\in\ENS{1}{\ndfe}^2$ by
\begin{equation}\label{eq:StiffElasElem}
\StiffElasElem_{i,j}(\simplex)=
\int_{\simplex} \Odv^t(\BaryCoorVec_j) \mathbb{C}\Odv(\BaryCoorVec_i)d\q,
\end{equation}
or equivalently, using~\eqref{def:BaryCoorVec}, we have for $1\leq\il,\jl\leq d+1$ and $1\leq l,n\leq m$
\begin{equation}\label{eq02:StiffElasElem}
\StiffElasElem_{i,j}(\simplex)=
\int_{\simplex} \Odv^t(\BaryCoorVec_{n,\beta}) \mathbb{C}\Odv(\BaryCoorVec_{l,\il})d\q,
\end{equation}
with $i=(\il-1)d+l$ and $j=(\jl-1)d+n$. The vectorization of $\StiffElasElem$ is based on the following result:
\begin{lemma}\label{lem:VecLocalElas}
  There exist two matrices $\MAT{Q}^{n,l}$ and $\MAT{S}^{n,l}$ of size $d$-by-$d$ depending only on $n$ and $l$ such that
 \begin{align}\label{eq:VecLocalElas}
\Odv^t(\BaryCoorVec_{n,\jl}) \MAT{C}\Odv(\BaryCoorVec_{l,\il})&=
\lambda \DOT{\GRAD\BaryCoor_\jl}{\MAT{Q}^{n,l} \GRAD\BaryCoor_\il}
+ \mu \DOT{\GRAD\BaryCoor_\jl}{ \MAT{S}^{n,l} \GRAD\BaryCoor_\il}.
\end{align}
\end{lemma}
The proof of Lemma~\ref{lem:VecLocalElas} is given in Appendix~\ref{app:ProofLem}.
\\
Using~\eqref{eq:VecLocalElas} in \eqref{eq02:StiffElasElem}, we have
\begin{align*}
\StiffElasElem_{i,j}(\simplex)
&=\DOT{\GRAD\BaryCoor_\jl}{\MAT{Q}^{n,l} \GRAD\BaryCoor_\il} \int_\simplex \lambda d\q
+\DOT{\GRAD\BaryCoor_\jl}{ \MAT{S}^{n,l} \GRAD\BaryCoor_\il} \int_\simplex \mu d\q.
\end{align*}
One possibility is to approximate the Lamé parameters $\lambda$ and $\mu$ by their $\Pk{1}$ finite element
interpolation $\PihK{1}(\lambda)$ and $\PihK{1}(\mu)$, respectively (we consider $\Pk{1}$ instead of $\Pk{0}$
to illustrate better the vectorization, the latter being a special case of the former).
Then we have
\begin{align}
\StiffElasElem_{i,j}(\simplex) & \approx
\frac{|K|}{d+1}\left(
\DOT{\GRAD\BaryCoor_\jl}{\MAT{Q}^{n,l} \GRAD\BaryCoor_\il} \lambda^s
+\DOT{\GRAD\BaryCoor_\jl}{ \MAT{S}^{n,l} \GRAD\BaryCoor_\il} \mu^s
\right),
\end{align}
with $\lambda^s=\sum_{\kl=1}^{d+1} \lambda(\q^\kl)$ and $\mu^s=\sum_{\kl=1}^{d+1} \mu(\q^\kl).$
The previous formula may now be vectorized as shown in Algorithm~\ref{algo:AssemblyStiffElasP1}.
This algorithm is based on the vectorization of the computation of the terms
$\DOT{\GRAD\BaryCoor_\jl}{\MAT{A} \GRAD\BaryCoor_\il}$, which is carried out with the
function \FName{dotMatVecG} in Algorithm~\ref{algo:dotMatVecG},
for any $d$-by-$d$ matrix $\MAT{A}$ independent of the
  $d$-\simplices of the mesh.
In this algorithm, $\MAT{G}$ is the array of gradients defined in~\eqref{eq:defGalpha},
$\il$ and $\jl$ are indices in $\ENS{1}{d+1}$, and
$\vecb{X}$ is a $1$-by-$\nme$ array such that
$\vecb{X}(k)=\DOT{\GRAD\BaryCoor_\jl}{\MAT{A} \GRAD\BaryCoor_\il}$ on $K=T_k.$
\begin{algorithm}[H]
\captionsetup{font=footnotesize}
\caption{Elastic stiffness matrix assembly - \texttt{\OptVec} version}
\label{algo:AssemblyStiffElasP1}
\ifdefined\IsInputOutput
\begin{minipage}{14.cm}
\footnotesize{\textbf{Input :}\\
  \begin{tabular}[t]{lcl}
     $\me$&:& $(d+1)$-by-$\nme$ connectivity array\\
     $\q$ & :& $\dd$-by-$\nq$ array of vertices coordinates\\
     $\volumes$ &: & $1$-by-$\nme$ array of volumes 
  \end{tabular}\\
\textbf{Output :}\\
  \begin{tabular}{lcl}
    $\MAT{M}$ & : & $\ndof$-by-$\ndof$ sparse matrix where $\ndof=d \nq.$
  \end{tabular}\\
}
\end{minipage}
\fi
\begin{algorithmic}[1]
\Function{$\MAT{M} \gets $\FNameDef{AssemblyStiffElasP1\OptVec}}{$\me,\q,\volumes,\Var{lamb},\Var{mu}$}
\State $[\MAT{Q},\MAT{S}] \gets \FName{MatQS}(d)$ \Comment{$\MAT{Q}$,$\MAT{S}$ : 2d array of matrices with
  $\MAT{Q}(l,n) = \MAT{Q}^{l,n}$} 
\State $\Var{Lambs}\ \gets \FNameStd{sum}(\Var{lamb}(\me),1)\MultVec\volumes/(d+1)$ 
\State $\Var{Mus} \gets \FNameStd{sum}(\Var{mu}(\me),1)\MultVec\volumes/(d+1)$
\State $\MAT{G}\gets \FName{GradientVec}(\q,\me)$ 
\State $\ndof \gets m*\nq,\ \ \MAT{M} \gets \MAT{O}_{\ndof}$ \Comment{$\ndof$-by-$\ndof$ sparse matrix}
\For{l}{1}{d}
  \For{\il}{1}{d+1}
    \State $ \vecb{I}_g \gets  m*(\me(\il,:)-1)+l$
    \For{n}{1}{d}
      \For{\jl}{1}{d+1}
	  \State $ \vecb{K}_g \gets \Var{Lambs}\MultVec\FName{dotMatVecG}(\MAT{Q}(l,n),\MAT{G},\il,\jl)$
	  \StatexIndent[6]        $+\Var{Mus}\MultVec\FName{dotMatVecG}(\MAT{S}(l,n),\MAT{G},\il,\jl)$
	  \State $ \vecb{J}_g \gets  m*(\me(\jl,:)-1)+n$
	  \State $\MAT{M} \gets \MAT{M} + \FNameStd{Sparse}(\vecb{I}_g,\vecb{J}_g,\vecb{K}_g,\ndof,\ndof)$
      \EndFor
    \EndFor
  \EndFor
\EndFor
\EndFunction
\end{algorithmic}
\end{algorithm}

\vspace{-4mm}
\begin{algorithm}[H]
\captionsetup{font=footnotesize}
\caption{Vectorization of $\vecb{X}$ in dimension $d$}
\label{algo:dotMatVecG}
\ifdefined\IsInputOutput
\Input{14.cm}{
  \begin{tabular}[t]{lcl}
    $\MAT{A}$   & : & $d$-by-$d$ matrix \\
    $\MAT{G}$ & : & gradients array ($\nme$-by-$(d+1)$-by-$d$)\\
    & & $\MAT{G}(k,\il,:)=\GRAD\BasisFunc_{\il}^k(\q),$ $\forall \il\in\ENS{1}{d+1}$\\
    $\il,\jl$ & : & index in $\ENS{1}{d+1}$
  \end{tabular}
}
\Output{14.cm}{
  \begin{tabular}{lcl}
    $\vecb{X}$ & : & $1$-by-$\nme$ array \\
    & & $\vecb{X}(k)=\DOT{\GRAD\BaryCoor_\jl}{\MAT{A} \GRAD\BaryCoor_\il}$ on $K=T_k.$
  \end{tabular}
}
\fi
\begin{algorithmic}[1]
\Function{$\footnotesize \vecb{X}\gets \FNameDef{dotMatVecG}$}{$\MAT{A},\MAT{G},\il,\jl$}
\State $\vecb{X} \gets \FNameStd{zeros}(1,\nme)$
\For{i}{1}{d}
  \For{j}{1}{d}
    \State $\vecb{X} \gets \vecb{X}+ \MAT{A}(j,i)*(\MAT{G}(:,\il,i)\MultVec\MAT{G}(:,\jl,j))$
  \EndFor
\EndFor
\EndFunction
\end{algorithmic}
\end{algorithm}

From Algorithm~\ref{algo:AssemblyStiffElasP1}, it is straightforward to derive
Algorithm~\ref{algo:AssemblyGenP1vectorOptV4s}
which uses the symmetry when the assembly matrix is symmetric.

We now present numerical results that illustrate the performance of
the finite element assembly methods presented in this article.
%
%

\section{Benchmark results}
\label{sec:NumRes}
We consider the assembly of the stiffness and elastic stiffness matrices in 2D and 3D,
in the following vector languages
\begin{itemize}
\item Matlab (R2014b),
\item Octave (3.8.1),
\item Python 3.4.0 with \textit{NumPy[1.8.2]} and \textit{SciPy[0.13.3]}.
\end{itemize}

We first compare the computation times of the different codes (\texttt{base},
\texttt{OptV1}, \texttt{OptV2}, \texttt{\OptVec} and \texttt{\OptVecS}), for each language considered.
Then we compare \texttt{\OptVecS} code with a C implementation of the assembly using
the \textit{SuiteSparse} library 4.2.1~\cite{Davis:SSP:2012} (``CXSparse'')
and with FreeFEM++.
A comparison of the performance of the \texttt{\OptVecS} code with recent and efficient Matlab/Octave codes
is also given. In every benchmark the domain $\Omega$ is the unit disk in 2D and the unit sphere in 3D.
For each result we present the average computation time for at least five finite element assembly calculations.

\subsection{Comparison of the \texttt{base}, \texttt{OptV1}, \texttt{OptV2}, \texttt{\OptVec} and \texttt{\OptVecS}  assembly codes}
\label{sub:CompVersions}
We show in Figures~\ref{fig:BenchStiff2D} and~\ref{fig:BenchStiffElas3D}, in logarithmic scales
and for each vector language, the performance of the assembly codes versus the matrix dimension $\ndof,$ for the 2D stiffness and 3D elastic stiffness matrices
    respectively.
We observe that the \texttt{\OptVecS} version is the fastest one and its complexity is $\bO{\ndof}$.

\BenchFigureNew{figure_StiffAssembling2DP1_Matlab}
            {figure_StiffAssembling2DP1_Octave}
            {figure_StiffAssembling2DP1_Python}
            {Stiffness matrix (2D): comparison of \texttt{base},  \texttt{OptV1},  \texttt{OptV2},  \texttt{\OptVec} and  \texttt{\OptVecS} codes in Matlab (top left),  Octave (top right) and Python (bottom).}
            {fig:BenchStiff2D}{htb}

\BenchFigureNew{figure_StiffElasAssembling3DP1_Matlab}
            {figure_StiffElasAssembling3DP1_Octave}
            {figure_StiffElasAssembling3DP1_Python}
            {Elastic stiffness matrix (3D): comparison of  \texttt{base},  \texttt{OptV1},  \texttt{OptV2},  \texttt{\OptVec} and  \texttt{\OptVecS} codes in Matlab (top left),  Octave (top right) and Python (bottom).}
            {fig:BenchStiffElas3D}{H}
\vspace{-0.7cm}

For the stiffness matrix in 2D, the \texttt{OptV1} version is about $40$, $95$ and $550$ times slower in Matlab, Python and Octave respectively.
Its numerical complexity is $\bO{\ndof}$.
The complexity of the less performing method, the \texttt{base} version,
is $\bO{\ndofs}$ in  Matlab and Octave, while it seems to be $\bO{\ndof}$ in Python.
This is partly due to the use of the  \texttt{LIL} format
in the sparse matrix assembly in Python, the conversion to the \texttt{CSC} format being included in
the computation time. We obtain similar results for the stiffness matrix in 3D and the elastic stiffness matrices
in 2D and 3D.
Computation times and \texttt{\OptVecS} speedup are given in Tables~\ref{tab:Stiff2D_Merge} and \ref{tab:StiffElas3D_Merge} for
the 2D stiffness and 3D elastic stiffness matrices respectively.
For the 3D stiffness and the 2D elastic stiffness matrices one can refer respectively to Tables~\ref{tab:Stiff3D_Merge} and~\ref{tab:StiffElas2D_Merge}.
We observe that the performance differences of the stiffness and elastic stiffness matrix assemblies
in 2D and 3D are partly due to the increase of the data: on the unit disk (2D) and the unit sphere (3D),
we have $\nme\approx 2 \nq$ and $\nme\approx 6\nq$ respectively.
For matrices of the same size (i.e. for an equal $\ndf$), in comparison to the 2D stiffness matrix, the number
of local values to be computed are 2, 4 and 16 times
larger for the 2D elastic stiffness, the 3D stiffness and the 3D elastic stiffness matrices
respectively.

\begin{table}[htbp]
\begin{center}
\noindent\adjustbox{max width=\textwidth}{\begin{tabular}{@{}c@{}}
\begin{tabular}{@{}cc@{}}
\textbf{StiffAssembling2DP1 - Matlab} & \textbf{StiffAssembling2DP1 - Octave} \\
\ifdefined\TabularWithCaption
\begin{table}[htbp]
\begin{center}
\noindent\adjustbox{max width=\textwidth}{
\fi
\begin{tabular}{@{}|r||*{5}{@{}c@{}|}@{}}
  \hline 
  $n_{dof}$ &  OptVS &  OptV &  OptV2 &  OptV1 &  base  \\ \hline \hline
$14222$ & \begin{tabular}{c} .063 {\tiny (s)}\\ \texttt{x 1} \end{tabular} & \begin{tabular}{c} .044 {\tiny (s)}\\ \texttt{x .692} \end{tabular} & \begin{tabular}{c} .053 {\tiny (s)}\\ \texttt{x .833} \end{tabular} & \begin{tabular}{c} 1.70 {\tiny (s)}\\ \texttt{x 26.8} \end{tabular} & \begin{tabular}{c} 6.79 {\tiny (s)}\\ \texttt{x 107} \end{tabular}\\ \hline
$125010$ & \begin{tabular}{c} .411 {\tiny (s)}\\ \texttt{x 1} \end{tabular} & \begin{tabular}{c} .617 {\tiny (s)}\\ \texttt{x 1.5} \end{tabular} & \begin{tabular}{c} .553 {\tiny (s)}\\ \texttt{x 1.35} \end{tabular} & \begin{tabular}{c} 14.4 {\tiny (s)}\\ \texttt{x 35.1} \end{tabular} & \begin{tabular}{c} 226 {\tiny (s)}\\ \texttt{x 550} \end{tabular}\\ \hline
$343082$ & \begin{tabular}{c} .985 {\tiny (s)}\\ \texttt{x 1} \end{tabular} & \begin{tabular}{c} 1.37 {\tiny (s)}\\ \texttt{x 1.39} \end{tabular} & \begin{tabular}{c} 1.36 {\tiny (s)}\\ \texttt{x 1.38} \end{tabular} & \begin{tabular}{c} 39.1 {\tiny (s)}\\ \texttt{x 39.7} \end{tabular} & \begin{tabular}{c} 1873 {\tiny (s)}\\ \texttt{x 1902} \end{tabular}\\ \hline
$885521$ & \begin{tabular}{c} 2.34 {\tiny (s)}\\ \texttt{x 1} \end{tabular} & \begin{tabular}{c} 3.24 {\tiny (s)}\\ \texttt{x 1.39} \end{tabular} & \begin{tabular}{c} 3.29 {\tiny (s)}\\ \texttt{x 1.41} \end{tabular} & \begin{tabular}{c} 99.7 {\tiny (s)}\\ \texttt{x 42.7} \end{tabular} & -\\ \hline
$1978602$ & \begin{tabular}{c} 5.45 {\tiny (s)}\\ \texttt{x 1} \end{tabular} & \begin{tabular}{c} 7.60 {\tiny (s)}\\ \texttt{x 1.40} \end{tabular} & \begin{tabular}{c} 7.28 {\tiny (s)}\\ \texttt{x 1.34} \end{tabular} & \begin{tabular}{c} 223 {\tiny (s)}\\ \texttt{x 40.9} \end{tabular} & -\\ \hline
\end{tabular}
\ifdefined\TabularWithCaption
}
\end{center}
\caption{OptV methods for Matlab : Stiff matrix in 2d}
\end{table}
\fi

&
\ifdefined\TabularWithCaption
\begin{table}[htbp]
\begin{center}
\noindent\adjustbox{max width=\textwidth}{
\fi
\begin{tabular}{@{}|r||*{5}{@{}c@{}|}@{}}
  \hline 
  $n_{dof}$ &  OptVS &  OptV &  OptV2 &  OptV1 &  base  \\ \hline \hline
$14222$ & \begin{tabular}{c} .017 {\tiny (s)}\\ \texttt{x 1} \end{tabular} & \begin{tabular}{c} .058 {\tiny (s)}\\ \texttt{x 3.36} \end{tabular} & \begin{tabular}{c} .036 {\tiny (s)}\\ \texttt{x 2.09} \end{tabular} & \begin{tabular}{c} 14.3 {\tiny (s)}\\ \texttt{x 826} \end{tabular} & \begin{tabular}{c} 15.4 {\tiny (s)}\\ \texttt{x 888} \end{tabular}\\ \hline
$125010$ & \begin{tabular}{c} .167 {\tiny (s)}\\ \texttt{x 1} \end{tabular} & \begin{tabular}{c} .218 {\tiny (s)}\\ \texttt{x 1.31} \end{tabular} & \begin{tabular}{c} .221 {\tiny (s)}\\ \texttt{x 1.33} \end{tabular} & \begin{tabular}{c} 124 {\tiny (s)}\\ \texttt{x 742} \end{tabular} & \begin{tabular}{c} 255 {\tiny (s)}\\ \texttt{x 1533} \end{tabular}\\ \hline
$343082$ & \begin{tabular}{c} .499 {\tiny (s)}\\ \texttt{x 1} \end{tabular} & \begin{tabular}{c} .656 {\tiny (s)}\\ \texttt{x 1.32} \end{tabular} & \begin{tabular}{c} .835 {\tiny (s)}\\ \texttt{x 1.67} \end{tabular} & \begin{tabular}{c} 340 {\tiny (s)}\\ \texttt{x 681} \end{tabular} & \begin{tabular}{c} 1458 {\tiny (s)}\\ \texttt{x 2923} \end{tabular}\\ \hline
$885521$ & \begin{tabular}{c} 1.47 {\tiny (s)}\\ \texttt{x 1} \end{tabular} & \begin{tabular}{c} 1.91 {\tiny (s)}\\ \texttt{x 1.30} \end{tabular} & \begin{tabular}{c} 2.43 {\tiny (s)}\\ \texttt{x 1.65} \end{tabular} & \begin{tabular}{c} 899 {\tiny (s)}\\ \texttt{x 613} \end{tabular} & -\\ \hline
$1978602$ & \begin{tabular}{c} 3.64 {\tiny (s)}\\ \texttt{x 1} \end{tabular} & \begin{tabular}{c} 4.63 {\tiny (s)}\\ \texttt{x 1.27} \end{tabular} & \begin{tabular}{c} 5.44 {\tiny (s)}\\ \texttt{x 1.49} \end{tabular} & \begin{tabular}{c} 2007 {\tiny (s)}\\ \texttt{x 551} \end{tabular} & -\\ \hline
\end{tabular}
\ifdefined\TabularWithCaption
}
\end{center}
\caption{OptV methods for Octave : Stiff matrix in 2d}
\end{table}
\fi

\end{tabular}
\vspace{0.15cm}\\

\begin{tabular}{@{}c@{}}
\textbf{StiffAssembling2DP1 - Python}\\
\ifdefined\TabularWithCaption
\begin{table}[htbp]
\begin{center}
\noindent\adjustbox{max width=\textwidth}{
\fi
\begin{tabular}{@{}|r||*{5}{@{}c@{}|}@{}}
  \hline 
  $n_{dof}$ &  OptVS &  OptV &  OptV2 &  OptV1 &  base  \\ \hline \hline
$14222$ & \begin{tabular}{c} .021 {\tiny (s)}\\ \texttt{x 1} \end{tabular} & \begin{tabular}{c} .027 {\tiny (s)}\\ \texttt{x 1.26} \end{tabular} & \begin{tabular}{c} .027 {\tiny (s)}\\ \texttt{x 1.29} \end{tabular} & \begin{tabular}{c} 2.64 {\tiny (s)}\\ \texttt{x 124} \end{tabular} & \begin{tabular}{c} 34.4 {\tiny (s)}\\ \texttt{x 1614} \end{tabular}\\ \hline
$125010$ & \begin{tabular}{c} .190 {\tiny (s)}\\ \texttt{x 1} \end{tabular} & \begin{tabular}{c} .241 {\tiny (s)}\\ \texttt{x 1.26} \end{tabular} & \begin{tabular}{c} .336 {\tiny (s)}\\ \texttt{x 1.77} \end{tabular} & \begin{tabular}{c} 23.2 {\tiny (s)}\\ \texttt{x 122} \end{tabular} & \begin{tabular}{c} 303 {\tiny (s)}\\ \texttt{x 1594} \end{tabular}\\ \hline
$343082$ & \begin{tabular}{c} .576 {\tiny (s)}\\ \texttt{x 1} \end{tabular} & \begin{tabular}{c} .716 {\tiny (s)}\\ \texttt{x 1.24} \end{tabular} & \begin{tabular}{c} .980 {\tiny (s)}\\ \texttt{x 1.70} \end{tabular} & \begin{tabular}{c} 63.5 {\tiny (s)}\\ \texttt{x 110} \end{tabular} & \begin{tabular}{c} 833 {\tiny (s)}\\ \texttt{x 1445} \end{tabular}\\ \hline
$885521$ & \begin{tabular}{c} 1.66 {\tiny (s)}\\ \texttt{x 1} \end{tabular} & \begin{tabular}{c} 2.05 {\tiny (s)}\\ \texttt{x 1.23} \end{tabular} & \begin{tabular}{c} 2.62 {\tiny (s)}\\ \texttt{x 1.58} \end{tabular} & \begin{tabular}{c} 164 {\tiny (s)}\\ \texttt{x 98.9} \end{tabular} & -\\ \hline
$1978602$ & \begin{tabular}{c} 3.92 {\tiny (s)}\\ \texttt{x 1} \end{tabular} & \begin{tabular}{c} 4.85 {\tiny (s)}\\ \texttt{x 1.24} \end{tabular} & \begin{tabular}{c} 6.04 {\tiny (s)}\\ \texttt{x 1.54} \end{tabular} & \begin{tabular}{c} 368 {\tiny (s)}\\ \texttt{x 93.9} \end{tabular} & -\\ \hline
\end{tabular}
\ifdefined\TabularWithCaption
}
\end{center}
\caption{OptV methods for Python : Stiff matrix in 2d}
\end{table}
\fi

\end{tabular}
\end{tabular}}
\vspace{0.15cm}

\captionof{table}{Stiffness matrix (2D) : comparison of \texttt{\OptVecS}, \texttt{\OptVec}, \texttt{OptV1} and \texttt{base} codes in Matlab (top left),
              Octave (top right) and Python (bottom) giving
              time in seconds (top value) and \texttt{\OptVecS} speedup (bottom value).}\label{tab:Stiff2D_Merge}
\end{center}
\end{table}

\begin{table}[htbp]
\begin{center}
\noindent\adjustbox{max width=\textwidth}{\begin{tabular}{@{}c@{}}
\begin{tabular}{@{}cc@{}}
\textbf{StiffElasAssembling3DP1 - Matlab} & \textbf{StiffElasAssembling3DP1 - Octave} \\
\ifdefined\TabularWithCaption
\begin{table}[htbp]
\begin{center}
\noindent\adjustbox{max width=\textwidth}{
\fi
\begin{tabular}{@{}|r||*{5}{@{}c@{}|}@{}}
  \hline 
  $n_{dof}$ &  OptVS &  OptV &  OptV2 &  OptV1 &  base  \\ \hline \hline
$16773$ & \begin{tabular}{c} .560 {\tiny (s)}\\ \texttt{x 1} \end{tabular} & \begin{tabular}{c} .971 {\tiny (s)}\\ \texttt{x 1.73} \end{tabular} & \begin{tabular}{c} .924 {\tiny (s)}\\ \texttt{x 1.65} \end{tabular} & \begin{tabular}{c} 67.6 {\tiny (s)}\\ \texttt{x 121} \end{tabular} & \begin{tabular}{c} 236 {\tiny (s)}\\ \texttt{x 422} \end{tabular}\\ \hline
$44124$ & \begin{tabular}{c} 1.70 {\tiny (s)}\\ \texttt{x 1} \end{tabular} & \begin{tabular}{c} 3.45 {\tiny (s)}\\ \texttt{x 2.03} \end{tabular} & \begin{tabular}{c} 2.60 {\tiny (s)}\\ \texttt{x 1.52} \end{tabular} & \begin{tabular}{c} 184 {\tiny (s)}\\ \texttt{x 108} \end{tabular} & \begin{tabular}{c} 1427 {\tiny (s)}\\ \texttt{x 837} \end{tabular}\\ \hline
$121710$ & \begin{tabular}{c} 4.43 {\tiny (s)}\\ \texttt{x 1} \end{tabular} & \begin{tabular}{c} 8.12 {\tiny (s)}\\ \texttt{x 1.83} \end{tabular} & \begin{tabular}{c} 7.55 {\tiny (s)}\\ \texttt{x 1.70} \end{tabular} & \begin{tabular}{c} 540 {\tiny (s)}\\ \texttt{x 122} \end{tabular} & \begin{tabular}{c} 1E+4 {\tiny (s)}\\ \texttt{x 2716} \end{tabular}\\ \hline
$601272$ & \begin{tabular}{c} 27.5 {\tiny (s)}\\ \texttt{x 1} \end{tabular} & \begin{tabular}{c} 47.4 {\tiny (s)}\\ \texttt{x 1.72} \end{tabular} & \begin{tabular}{c} 41.5 {\tiny (s)}\\ \texttt{x 1.51} \end{tabular} & \begin{tabular}{c} 2765 {\tiny (s)}\\ \texttt{x 101} \end{tabular} & -\\ \hline
$1144680$ & \begin{tabular}{c} 51.5 {\tiny (s)}\\ \texttt{x 1} \end{tabular} & \begin{tabular}{c} 89.4 {\tiny (s)}\\ \texttt{x 1.74} \end{tabular} & \begin{tabular}{c} 84.2 {\tiny (s)}\\ \texttt{x 1.64} \end{tabular} & \begin{tabular}{c} 5254 {\tiny (s)}\\ \texttt{x 102} \end{tabular} & -\\ \hline
\end{tabular}
\ifdefined\TabularWithCaption
}
\end{center}
\caption{OptV methods for Matlab : StiffElas matrix in 3d}
\end{table}
\fi

&
\ifdefined\TabularWithCaption
\begin{table}[htbp]
\begin{center}
\noindent\adjustbox{max width=\textwidth}{
\fi
\begin{tabular}{@{}|r||*{5}{@{}c@{}|}@{}}
  \hline 
  $n_{dof}$ &  OptVS &  OptV &  OptV2 &  OptV1 &  base  \\ \hline \hline
$16773$ & \begin{tabular}{c} .364 {\tiny (s)}\\ \texttt{x 1} \end{tabular} & \begin{tabular}{c} .628 {\tiny (s)}\\ \texttt{x 1.73} \end{tabular} & \begin{tabular}{c} .569 {\tiny (s)}\\ \texttt{x 1.56} \end{tabular} & \begin{tabular}{c} 255 {\tiny (s)}\\ \texttt{x 701} \end{tabular} & \begin{tabular}{c} 321 {\tiny (s)}\\ \texttt{x 882} \end{tabular}\\ \hline
$44124$ & \begin{tabular}{c} .993 {\tiny (s)}\\ \texttt{x 1} \end{tabular} & \begin{tabular}{c} 1.69 {\tiny (s)}\\ \texttt{x 1.71} \end{tabular} & \begin{tabular}{c} 1.49 {\tiny (s)}\\ \texttt{x 1.50} \end{tabular} & \begin{tabular}{c} 698 {\tiny (s)}\\ \texttt{x 703} \end{tabular} & \begin{tabular}{c} 1314 {\tiny (s)}\\ \texttt{x 1323} \end{tabular}\\ \hline
$121710$ & \begin{tabular}{c} 3.03 {\tiny (s)}\\ \texttt{x 1} \end{tabular} & \begin{tabular}{c} 5.13 {\tiny (s)}\\ \texttt{x 1.69} \end{tabular} & \begin{tabular}{c} 4.19 {\tiny (s)}\\ \texttt{x 1.38} \end{tabular} & \begin{tabular}{c} 1976 {\tiny (s)}\\ \texttt{x 651} \end{tabular} & \begin{tabular}{c} 9338 {\tiny (s)}\\ \texttt{x 3078} \end{tabular}\\ \hline
$601272$ & \begin{tabular}{c} 18.9 {\tiny (s)}\\ \texttt{x 1} \end{tabular} & \begin{tabular}{c} 31.7 {\tiny (s)}\\ \texttt{x 1.68} \end{tabular} & \begin{tabular}{c} 25.5 {\tiny (s)}\\ \texttt{x 1.35} \end{tabular} & \begin{tabular}{c} 9853 {\tiny (s)}\\ \texttt{x 521} \end{tabular} & -\\ \hline
$1144680$ & \begin{tabular}{c} 40.9 {\tiny (s)}\\ \texttt{x 1} \end{tabular} & \begin{tabular}{c} 69.1 {\tiny (s)}\\ \texttt{x 1.69} \end{tabular} & \begin{tabular}{c} 55.6 {\tiny (s)}\\ \texttt{x 1.36} \end{tabular} & \begin{tabular}{c} 2E+4 {\tiny (s)}\\ \texttt{x 471} \end{tabular} & -\\ \hline
\end{tabular}
\ifdefined\TabularWithCaption
}
\end{center}
\caption{OptV methods for Octave : StiffElas matrix in 3d}
\end{table}
\fi

\end{tabular}
\vspace{0.15cm}\\

\begin{tabular}{@{}c@{}}
\textbf{StiffElasAssembling3DP1 - Python}\\
\ifdefined\TabularWithCaption
\begin{table}[htbp]
\begin{center}
\noindent\adjustbox{max width=\textwidth}{
\fi
\begin{tabular}{@{}|r||*{5}{@{}c@{}|}@{}}
  \hline 
  $n_{dof}$ &  OptVS &  OptV &  OptV2 &  OptV1 &  base  \\ \hline \hline
$16773$ & \begin{tabular}{c} .391 {\tiny (s)}\\ \texttt{x 1} \end{tabular} & \begin{tabular}{c} .622 {\tiny (s)}\\ \texttt{x 1.59} \end{tabular} & \begin{tabular}{c} .486 {\tiny (s)}\\ \texttt{x 1.24} \end{tabular} & \begin{tabular}{c} 122 {\tiny (s)}\\ \texttt{x 312} \end{tabular} & \begin{tabular}{c} 784 {\tiny (s)}\\ \texttt{x 2004} \end{tabular}\\ \hline
$44124$ & \begin{tabular}{c} .954 {\tiny (s)}\\ \texttt{x 1} \end{tabular} & \begin{tabular}{c} 1.56 {\tiny (s)}\\ \texttt{x 1.63} \end{tabular} & \begin{tabular}{c} 1.32 {\tiny (s)}\\ \texttt{x 1.38} \end{tabular} & \begin{tabular}{c} 333 {\tiny (s)}\\ \texttt{x 349} \end{tabular} & \begin{tabular}{c} 2141 {\tiny (s)}\\ \texttt{x 2243} \end{tabular}\\ \hline
$121710$ & \begin{tabular}{c} 2.55 {\tiny (s)}\\ \texttt{x 1} \end{tabular} & \begin{tabular}{c} 4.21 {\tiny (s)}\\ \texttt{x 1.65} \end{tabular} & \begin{tabular}{c} 3.79 {\tiny (s)}\\ \texttt{x 1.49} \end{tabular} & \begin{tabular}{c} 946 {\tiny (s)}\\ \texttt{x 372} \end{tabular} & \begin{tabular}{c} 6071 {\tiny (s)}\\ \texttt{x 2384} \end{tabular}\\ \hline
$601272$ & \begin{tabular}{c} 16.4 {\tiny (s)}\\ \texttt{x 1} \end{tabular} & \begin{tabular}{c} 27.6 {\tiny (s)}\\ \texttt{x 1.68} \end{tabular} & \begin{tabular}{c} 24.9 {\tiny (s)}\\ \texttt{x 1.52} \end{tabular} & \begin{tabular}{c} 4850 {\tiny (s)}\\ \texttt{x 296} \end{tabular} & -\\ \hline
$1144680$ & \begin{tabular}{c} 36.4 {\tiny (s)}\\ \texttt{x 1} \end{tabular} & \begin{tabular}{c} 61.5 {\tiny (s)}\\ \texttt{x 1.69} \end{tabular} & \begin{tabular}{c} 54.2 {\tiny (s)}\\ \texttt{x 1.49} \end{tabular} & - & -\\ \hline
\end{tabular}
\ifdefined\TabularWithCaption
}
\end{center}
\caption{OptV methods for Python : StiffElas matrix in 3d}
\end{table}
\fi

\end{tabular}
\end{tabular}}
\vspace{0.15cm}

\captionof{table}{Elastic stiffness matrix (3D) : comparison of \texttt{\OptVecS}, \texttt{\OptVec}, \texttt{OptV2}, \texttt{OptV1} and \texttt{base} codes in Matlab (top left),
              Octave (top right) and Python (bottom) giving
              time in seconds (top value) and \texttt{\OptVecS}  speedup (bottom value).}\label{tab:StiffElas3D_Merge}
\end{center}
\end{table}

\begin{figure}[H]
  \centering
  \includegraphics*[scale={0.72}]{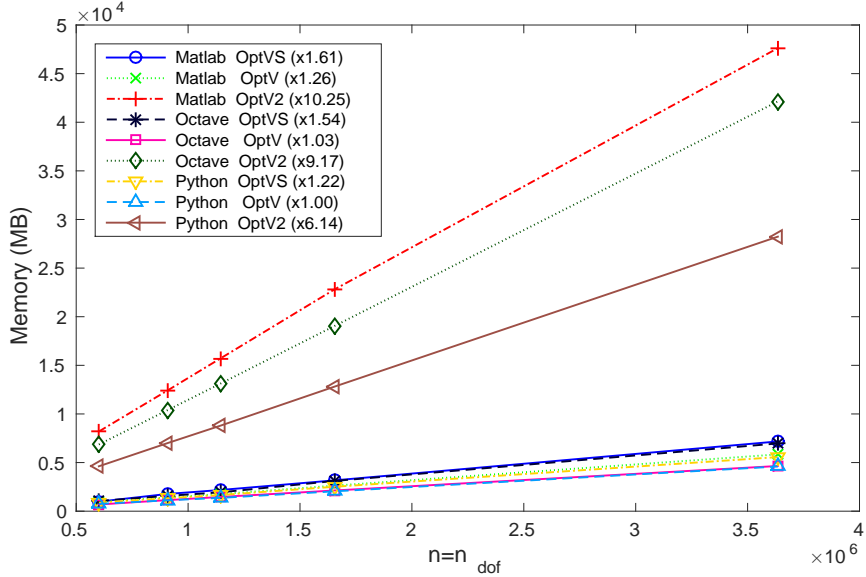}
  \caption{Elastic stiffness matrix (3D): memory usage in MB\label{fig:CompStiffElasAssembling3DP1mem}
   and ratio between the slope of each method and \texttt{\OptVec} in Python (in the caption)}
\end{figure}

In Figure~\ref{fig:CompStiffElasAssembling3DP1mem} we compare the maximum of memory for \texttt{\OptVecS}, \texttt{\OptVec} and \texttt{OptV2} codes.
The \texttt{OptV2} method is more consuming than \texttt{\OptVecS} and \texttt{\OptVec} respectively
by a factor between 5 and 6.3 and between 6 and 8.9 depending on the language.

\subsection{Comparison of the \texttt{\OptVecS} version with CXSparse and FreeFEM++}

In Tables~\ref{table:CompStiff2DandStiffElas3DAssembling}
the \texttt{\OptVecS} codes in Matlab/Octave/Python are compared
with a C implementation of the assembly (\texttt{OptV1} version) using
the \textit{SuiteSparse} library~\cite{Davis:SSP:2012} (``CXSparse'')
and with a FreeFEM++ code
for the stiffness matrix in 2D and the elastic stiffness matrix in 3D.

The computation cost for the
stiffness matrix in 3D and the elastic stiffness matrix in 2D are given in
Tables~\ref{table:CompStiffAssembling3DP1} and \ref{table:CompStiffElasAssembling2DP1}.
We observe that \texttt{\OptVecS} version is approximately 1.5 and 5.5 times in Matlab, 2 and 7.5 times in Octave, and 2.1 and 8.5 times in Python faster than FreeFEM++.
Compared to C,
computation times are multiplied by a factor between 2.5 and 4.7 in Matlab,
1.9 and 3.7 in Octave, and 1.8 and 3.2 in Python. Unlike what is commonly believed the performance is not radically worse than that of C.

\let\TabularWithCaption\undefined
\begin{table}[htbp]
\begin{center}
\noindent\adjustbox{width=10.cm}{
\ifdefined\TabularWithCaption
\begin{table}[htbp]
\begin{center}
\noindent\adjustbox{max width=\textwidth}{
\fi
\begin{tabular}{@{}|r||*{5}{@{}c@{}|}@{}}
  \hline 
  $n_{dof}$ &  \begin{tabular}{c} CXSparse\\ (4.2.1) \end{tabular} &  \begin{tabular}{c} Matlab\\ (2014b) \end{tabular} &  \begin{tabular}{c} Octave\\ (3.8.1) \end{tabular} &  \begin{tabular}{c} Python\\ (3.4.0) \end{tabular} &  \begin{tabular}{c} FreeFEM\\ (3.31) \end{tabular}  \\ \hline \hline
$14222$ & \begin{tabular}{c} 0.014 (s)\\ \texttt{x 1.00} \end{tabular} & \begin{tabular}{c} 0.063 (s)\\ \texttt{x 4.64} \end{tabular} & \begin{tabular}{c} 0.017 (s)\\ \texttt{x 1.26} \end{tabular} & \begin{tabular}{c} 0.021 (s)\\ \texttt{x 1.56} \end{tabular} & \begin{tabular}{c} 0.071 (s)\\ \texttt{x 5.16} \end{tabular}\\ \hline
$125010$ & \begin{tabular}{c} 0.073 (s)\\ \texttt{x 1.00} \end{tabular} & \begin{tabular}{c} 0.411 (s)\\ \texttt{x 5.62} \end{tabular} & \begin{tabular}{c} 0.167 (s)\\ \texttt{x 2.28} \end{tabular} & \begin{tabular}{c} 0.190 (s)\\ \texttt{x 2.60} \end{tabular} & \begin{tabular}{c} 0.501 (s)\\ \texttt{x 6.85} \end{tabular}\\ \hline
$343082$ & \begin{tabular}{c} 0.221 (s)\\ \texttt{x 1.00} \end{tabular} & \begin{tabular}{c} 0.985 (s)\\ \texttt{x 4.46} \end{tabular} & \begin{tabular}{c} 0.499 (s)\\ \texttt{x 2.26} \end{tabular} & \begin{tabular}{c} 0.576 (s)\\ \texttt{x 2.61} \end{tabular} & \begin{tabular}{c} 1.421 (s)\\ \texttt{x 6.43} \end{tabular}\\ \hline
$885521$ & \begin{tabular}{c} 0.606 (s)\\ \texttt{x 1.00} \end{tabular} & \begin{tabular}{c} 2.337 (s)\\ \texttt{x 3.86} \end{tabular} & \begin{tabular}{c} 1.467 (s)\\ \texttt{x 2.42} \end{tabular} & \begin{tabular}{c} 1.660 (s)\\ \texttt{x 2.74} \end{tabular} & \begin{tabular}{c} 3.692 (s)\\ \texttt{x 6.10} \end{tabular}\\ \hline
$1978602$ & \begin{tabular}{c} 1.354 (s)\\ \texttt{x 1.00} \end{tabular} & \begin{tabular}{c} 5.446 (s)\\ \texttt{x 4.02} \end{tabular} & \begin{tabular}{c} 3.644 (s)\\ \texttt{x 2.69} \end{tabular} & \begin{tabular}{c} 3.920 (s)\\ \texttt{x 2.89} \end{tabular} & \begin{tabular}{c} 8.305 (s)\\ \texttt{x 6.13} \end{tabular}\\ \hline
\end{tabular}
\ifdefined\TabularWithCaption
}
\end{center}
\caption{2D stiffness matrix : computational cost versus $n_{dof},$ with the \texttt{\OptVecS} Matlab/Octave/Python version ($2^{\mbox{nd}}/3^{\mbox{rd}}/4^{\mbox{th}}$ columns), with CXSparse ($1^{\mbox{st}}$ column) and FreeFEM++ ($5^{\mbox{th}}$ column) : time in seconds (top value) and speedup (bottom value). The speedup reference is CXSparse code.\label{table:CompStiffAssembling2DP1}}
\end{table}
\fi

}\vspace{3mm}
\noindent\adjustbox{width=10.cm}{
\ifdefined\TabularWithCaption
\begin{table}[htbp]
\begin{center}
\noindent\adjustbox{max width=\textwidth}{
\fi
\begin{tabular}{@{}|r||*{5}{@{}c@{}|}@{}}
  \hline 
  $n_{dof}$ &  \begin{tabular}{c} CXSparse\\ (4.2.1) \end{tabular} &  \begin{tabular}{c} Matlab\\ (2014b) \end{tabular} &  \begin{tabular}{c} Octave\\ (3.8.1) \end{tabular} &  \begin{tabular}{c} Python\\ (3.4.0) \end{tabular} &  \begin{tabular}{c} FreeFEM\\ (3.31) \end{tabular}  \\ \hline \hline
$16773$ & \begin{tabular}{c} 0.137 (s)\\ \texttt{x 1.00} \end{tabular} & \begin{tabular}{c} 0.560 (s)\\ \texttt{x 4.09} \end{tabular} & \begin{tabular}{c} 0.364 (s)\\ \texttt{x 2.66} \end{tabular} & \begin{tabular}{c} 0.391 (s)\\ \texttt{x 2.86} \end{tabular} & \begin{tabular}{c} 3.827 (s)\\ \texttt{x 27.95} \end{tabular}\\ \hline
$44124$ & \begin{tabular}{c} 0.398 (s)\\ \texttt{x 1.00} \end{tabular} & \begin{tabular}{c} 1.705 (s)\\ \texttt{x 4.29} \end{tabular} & \begin{tabular}{c} 0.993 (s)\\ \texttt{x 2.50} \end{tabular} & \begin{tabular}{c} 0.954 (s)\\ \texttt{x 2.40} \end{tabular} & \begin{tabular}{c} 10.440 (s)\\ \texttt{x 26.26} \end{tabular}\\ \hline
$121710$ & \begin{tabular}{c} 1.193 (s)\\ \texttt{x 1.00} \end{tabular} & \begin{tabular}{c} 4.433 (s)\\ \texttt{x 3.72} \end{tabular} & \begin{tabular}{c} 3.034 (s)\\ \texttt{x 2.54} \end{tabular} & \begin{tabular}{c} 2.547 (s)\\ \texttt{x 2.13} \end{tabular} & \begin{tabular}{c} 29.914 (s)\\ \texttt{x 25.08} \end{tabular}\\ \hline
$601272$ & \begin{tabular}{c} 6.386 (s)\\ \texttt{x 1.00} \end{tabular} & \begin{tabular}{c} 27.482 (s)\\ \texttt{x 4.30} \end{tabular} & \begin{tabular}{c} 18.894 (s)\\ \texttt{x 2.96} \end{tabular} & \begin{tabular}{c} 16.359 (s)\\ \texttt{x 2.56} \end{tabular} & \begin{tabular}{c} 152.553 (s)\\ \texttt{x 23.89} \end{tabular}\\ \hline
$1144680$ & \begin{tabular}{c} 12.477 (s)\\ \texttt{x 1.00} \end{tabular} & \begin{tabular}{c} 51.465 (s)\\ \texttt{x 4.12} \end{tabular} & \begin{tabular}{c} 40.940 (s)\\ \texttt{x 3.28} \end{tabular} & \begin{tabular}{c} 36.392 (s)\\ \texttt{x 2.92} \end{tabular} & \begin{tabular}{c} 293.307 (s)\\ \texttt{x 23.51} \end{tabular}\\ \hline
\end{tabular}
\ifdefined\TabularWithCaption
}
\end{center}
\caption{3D elastic stiffness matrix : computational cost versus $n_{dof},$ with the \texttt{\OptVecS} Matlab/Octave/Python version ($2^{\mbox{nd}}/3^{\mbox{rd}}/4^{\mbox{th}}$ columns), with CXSparse ($1^{\mbox{st}}$ column) and FreeFEM++ ($5^{\mbox{th}}$ column) : time in seconds (top value) and speedup (bottom value). The speedup reference is CXSparse code.\label{table:CompStiffElasAssembling3DP1}}
\end{table}
\fi

}
\end{center}
\caption{2D Stiffness matrix (top table) and 3D elastic stiffness matrix (bottom table) :
         computational cost versus $n_{dof},$ with the \texttt{\OptVecS}
         Matlab/Octave/Python version ($2^{\mbox{nd}}/3^{\mbox{rd}}/4^{\mbox{th}}$ columns),
         with CXSparse ($1^{\mbox{st}}$ column) and FreeFEM++ ($5^{\mbox{th}}$ column);
         time in seconds (top value) and CXSparse speedup (bottom value).
         \label{table:CompStiff2DandStiffElas3DAssembling}}
\end{table}

\subsection{Comparison with other matrix assemblies in Matlab and Octave}
In Matlab/Octave other efficient algorithms have been proposed recently
in~\cite{Chen:PFE:2011,Chen:iFEM:2013,Hannukainen:IFE:2012,Rahman:FMA:2011}.
More precisely, in~\cite{Hannukainen:IFE:2012}, a vectorization is proposed, based on the permutation
of two local loops with the one through the elements. This technique
allows to easily assemble different matrices, from  a reference element by affine transformation
and by using a numerical integration.
In~\cite{Rahman:FMA:2011}, the implementation is based on extending element operations on arrays
into operations on arrays of matrices, calling them matrix-array operations. The array elements are matrices
instead of scalars and the operations are defined by the rules of linear algebra.
Thanks to these new tools and a quadrature formula, different matrices are computed
without any loop.
In~\cite{Chen:iFEM:2013},
for the assembly of the stiffness matrix in 2D associated to $\Pk{1}$
  finite elements,
L.~Chen constructs vectorially the nine sparse matrices corresponding to the
nine elements of the local stiffness matrix in 2D and adds them to obtain the global matrix.
The restriction to $d=2$ or $3$ of Algorithm~\ref{algo:AssemblyGenP1OptVec} corresponds to the method
in~\cite{Chen:iFEM:2013}.

We compare these codes to \texttt{\OptVecS} for the assembly of the stiffness matrix in~2D.
In Tables~\ref{StiffMatAll} and~\ref{StiffOctAll}, using Matlab and Octave respectively,
computation times versus the number of vertices are given for the different codes.
\texttt{\OptVecS} speedup is between $1$ and $2.5$  in comparison with the other vectorized codes for sufficiently
  fine meshes.
\def\TabularWithCaption{}
\ifdefined\TabularWithCaption
\begin{table}[htbp]
\begin{center}
\noindent\adjustbox{max width=\textwidth}{
\fi
\begin{tabular}{@{}|r||*{5}{@{}c@{}|}@{}}
  \hline 
  $n_{dof}$ &  OptVs &  Chen &  iFEM &  HanJun &  RahVal  \\ \hline \hline
$125010$ & \begin{tabular}{c} 0.411 (s)\\ \texttt{x 1.00} \end{tabular} & \begin{tabular}{c} 0.616 (s)\\ \texttt{x 1.50} \end{tabular} & \begin{tabular}{c} 0.693 (s)\\ \texttt{x 1.69} \end{tabular} & \begin{tabular}{c} 0.646 (s)\\ \texttt{x 1.57} \end{tabular} & \begin{tabular}{c} 0.664 (s)\\ \texttt{x 1.61} \end{tabular}\\ \hline
$343082$ & \begin{tabular}{c} 0.985 (s)\\ \texttt{x 1.00} \end{tabular} & \begin{tabular}{c} 1.464 (s)\\ \texttt{x 1.49} \end{tabular} & \begin{tabular}{c} 1.257 (s)\\ \texttt{x 1.28} \end{tabular} & \begin{tabular}{c} 1.989 (s)\\ \texttt{x 2.02} \end{tabular} & \begin{tabular}{c} 2.096 (s)\\ \texttt{x 2.13} \end{tabular}\\ \hline
$885521$ & \begin{tabular}{c} 2.337 (s)\\ \texttt{x 1.00} \end{tabular} & \begin{tabular}{c} 3.307 (s)\\ \texttt{x 1.41} \end{tabular} & \begin{tabular}{c} 2.966 (s)\\ \texttt{x 1.27} \end{tabular} & \begin{tabular}{c} 4.372 (s)\\ \texttt{x 1.87} \end{tabular} & \begin{tabular}{c} 4.721 (s)\\ \texttt{x 2.02} \end{tabular}\\ \hline
$1978602$ & \begin{tabular}{c} 5.446 (s)\\ \texttt{x 1.00} \end{tabular} & \begin{tabular}{c} 9.286 (s)\\ \texttt{x 1.71} \end{tabular} & \begin{tabular}{c} 7.221 (s)\\ \texttt{x 1.33} \end{tabular} & \begin{tabular}{c} 9.813 (s)\\ \texttt{x 1.80} \end{tabular} & \begin{tabular}{c} 9.123 (s)\\ \texttt{x 1.68} \end{tabular}\\ \hline
$3085628$ & \begin{tabular}{c} 8.644 (s)\\ \texttt{x 1.00} \end{tabular} & \begin{tabular}{c} 12.332 (s)\\ \texttt{x 1.43} \end{tabular} & \begin{tabular}{c} 11.444 (s)\\ \texttt{x 1.32} \end{tabular} & \begin{tabular}{c} 14.562 (s)\\ \texttt{x 1.68} \end{tabular} & \begin{tabular}{c} 14.841 (s)\\ \texttt{x 1.72} \end{tabular}\\ \hline
\end{tabular}
\ifdefined\TabularWithCaption
}
\end{center}
\caption{Stiffness matrix (2D): computational cost in Matlab (R2014b) versus $\nq,$ with the \texttt{\OptVecS} version (column $2$) and with the codes in \cite{Chen:PFE:2011,Chen:iFEM:2013,Hannukainen:IFE:2012,Rahman:FMA:2011} (columns $3$-$6$) : time in seconds (top value) and speedup (bottom value). The speedup reference is \texttt{\OptVecS} version.\label{StiffMatAll}}
\end{table}
\fi

\vspace{3mm}
\ifdefined\TabularWithCaption
\begin{table}[htbp]
\begin{center}
\noindent\adjustbox{max width=\textwidth}{
\fi
\begin{tabular}{@{}|r||*{5}{@{}c@{}|}@{}}
  \hline 
  $n_{dof}$ &  OptVs &  Chen &  iFEM &  HanJun &  RahVal  \\ \hline \hline
$125010$ & \begin{tabular}{c} 0.167 (s)\\ \texttt{x 1.00} \end{tabular} & \begin{tabular}{c} 0.305 (s)\\ \texttt{x 1.83} \end{tabular} & \begin{tabular}{c} 0.288 (s)\\ \texttt{x 1.73} \end{tabular} & \begin{tabular}{c} 0.417 (s)\\ \texttt{x 2.50} \end{tabular} & \begin{tabular}{c} 0.486 (s)\\ \texttt{x 2.92} \end{tabular}\\ \hline
$343082$ & \begin{tabular}{c} 0.499 (s)\\ \texttt{x 1.00} \end{tabular} & \begin{tabular}{c} 0.823 (s)\\ \texttt{x 1.65} \end{tabular} & \begin{tabular}{c} 0.644 (s)\\ \texttt{x 1.29} \end{tabular} & \begin{tabular}{c} 1.299 (s)\\ \texttt{x 2.60} \end{tabular} & \begin{tabular}{c} 1.245 (s)\\ \texttt{x 2.50} \end{tabular}\\ \hline
$885521$ & \begin{tabular}{c} 1.467 (s)\\ \texttt{x 1.00} \end{tabular} & \begin{tabular}{c} 2.123 (s)\\ \texttt{x 1.45} \end{tabular} & \begin{tabular}{c} 1.663 (s)\\ \texttt{x 1.13} \end{tabular} & \begin{tabular}{c} 3.720 (s)\\ \texttt{x 2.54} \end{tabular} & \begin{tabular}{c} 3.221 (s)\\ \texttt{x 2.20} \end{tabular}\\ \hline
$1978602$ & \begin{tabular}{c} 3.644 (s)\\ \texttt{x 1.00} \end{tabular} & \begin{tabular}{c} 4.674 (s)\\ \texttt{x 1.28} \end{tabular} & \begin{tabular}{c} 3.832 (s)\\ \texttt{x 1.05} \end{tabular} & \begin{tabular}{c} 8.279 (s)\\ \texttt{x 2.27} \end{tabular} & \begin{tabular}{c} 7.164 (s)\\ \texttt{x 1.97} \end{tabular}\\ \hline
$3085628$ & \begin{tabular}{c} 6.457 (s)\\ \texttt{x 1.00} \end{tabular} & \begin{tabular}{c} 7.786 (s)\\ \texttt{x 1.21} \end{tabular} & \begin{tabular}{c} 6.642 (s)\\ \texttt{x 1.03} \end{tabular} & \begin{tabular}{c} 13.523 (s)\\ \texttt{x 2.09} \end{tabular} & \begin{tabular}{c} 11.583 (s)\\ \texttt{x 1.79} \end{tabular}\\ \hline
\end{tabular}
\ifdefined\TabularWithCaption
}
\end{center}
\caption{Stiffness matrix (2D): computational cost in Octave (3.8.1) versus $\nq,$ with the \texttt{\OptVecS} version (column $2$) and with the codes in \cite{Chen:PFE:2011,Chen:iFEM:2013,Hannukainen:IFE:2012,Rahman:FMA:2011} (columns $3$-$6$) : time in seconds (top value) and speedup (bottom value). The speedup reference is \texttt{\OptVecS} version.\label{StiffOctAll}}
\end{table}
\fi

\begin{figure}[H]
  \centering
  \includegraphics*[scale={0.45}]{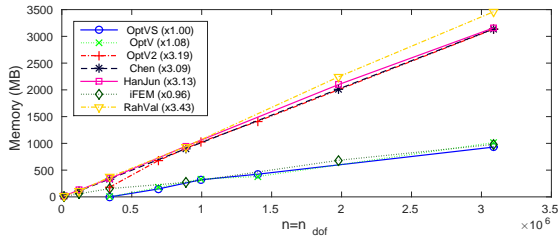}
  \caption{Stiffness matrix (2D): memory usage in MB and ratio between the slope of each method and \texttt{\OptVecS} (in the caption) \label{fig:CompStiffAssembling2DP1mem}}
\end{figure}

In Figure~\ref{fig:CompStiffAssembling2DP1mem} we compare the memory costs in Matlab of our assembly codes with
the other ones. As expected the consumption of \texttt{\OptVecS} and \texttt{\OptVec} methods are observed
  to be close to that of \texttt{iFEM} and lower than that of the other codes.

\vspace{-2mm}

\section{Conclusion and work in progress}
We presented vectorized algorithms for the assembly of $\Pk{1}$ finite element matrices in arbitrary dimension.
The implementation of these algorithms has been done in different vector languages such as Matlab, Octave and Python
to calculate the stiffness and elastic stiffness matrices.
Computation times of different versions (vectorized or not) have been compared in several interpreted languages and C.
Numerical examples show the efficiency of the \texttt{OptV2}, \texttt{\OptVec} and \texttt{\OptVecS} algorithms.
More precisely, for the \texttt{\OptVecS} method,
the assembly of the stiffness matrix in 2D of size $10^6$ is performed
in $2.6$, $1.75$ and $2$ seconds with Matlab, Octave and Python respectively
and in $0.75$ seconds with~C.
Less performance is obtained for the assembly of the elastic stiffness matrix in 3D: a matrix of size $10^6$
is computed in $45$, $35.8$ and $31.8$ seconds, with Matlab, Octave and Python respectively
and in $10.9$ seconds with~C.
Moreover we observed that \texttt{\OptVecS} is about 1.4 times faster
than the non-symmetrized versions \texttt{\OptVec} and \texttt{OptV2}.
\texttt{\OptVec} and \texttt{\OptVecS} methods are less memory consuming than \texttt{OptV2}.
Preliminary results towards the extension to
 $\Pk{k}$ finite elements are given in the Appendix.
The algorithms in arbitrary dimension for piecewise polynomials of higher order,
is the subject of a future paper.
The \texttt{OptV2} algorithm has been also implemented with a NVIDIA GPU\footnote{GeForce GTX Titan Black, 2880 CUDA Core, 6Go Memory}, using
 the \textit{Thrust} and \textit{Cusp} libraries. For the 2D elastic stiffness and 3D stiffness matrices,
  the \texttt{OptV2} code is respectively 3.5 and 7 times faster on GPU than the C code
 (the time for GPU/CPU data and matrix transfers is taken into account).

Vectorization gave good performance and the vectorized code can be used for other matrices or discretizations,
the only part of the code that have to be reviewed (which is probably the most difficult part) is the vectorization of
the element matrix computation.
We have seen that it is possible to efficiently assemble matrices of large size in interpreted languages.
  In this framework  Python showed some very good performance even though Octave seems to be more efficient
  in some cases. Moreover the performance of our vectorized codes was better in Octave than in Matlab.
  The Python and Matlab/Octave codes are available online (see \cite{CJS:Software:2015}).

\appendix

\section{Additional benchmark results}\label{appendix:Comp}
In this section, we consider the assembly of the 3D stiffness and 2D elastic stiffness matrices.
In Tables~\ref{table:CompStiffAssembling3DP1} and~\ref{table:CompStiffElasAssembling2DP1}
we compare the \texttt{\OptVecS} versions in Matlab/Octave/Python
with a~C implementation of the assembly (\texttt{OptV1} version) using
the \textit{SuiteSparse} library~\cite{Davis:SSP:2012} (``CXSparse''),
and with a FreeFEM++ version.
In Tables~\ref{tab:Stiff3D_Merge} and~\ref{tab:StiffElas2D_Merge}
the computation times of \texttt{\OptVecS}, \texttt{\OptVec}, \texttt{OptV2}, \texttt{OptV1} and \texttt{base} versions are compared in Matlab, Octave and Python. We observe similar results as in Section~\ref{sec:NumRes}.

\def\TabularWithCaption{}
\ifdefined\TabularWithCaption
\begin{table}[htbp]
\begin{center}
\noindent\adjustbox{max width=\textwidth}{
\fi
\begin{tabular}{@{}|r||*{5}{@{}c@{}|}@{}}
  \hline 
  $n_{dof}$ &  \begin{tabular}{c} CXSparse\\ (4.2.1) \end{tabular} &  \begin{tabular}{c} Matlab\\ (2014b) \end{tabular} &  \begin{tabular}{c} Octave\\ (3.8.1) \end{tabular} &  \begin{tabular}{c} Python\\ (3.4.0) \end{tabular} &  \begin{tabular}{c} FreeFEM\\ (3.31) \end{tabular}  \\ \hline \hline
$14708$ & \begin{tabular}{c} 0.087 (s)\\ \texttt{x 1.00} \end{tabular} & \begin{tabular}{c} 0.135 (s)\\ \texttt{x 1.55} \end{tabular} & \begin{tabular}{c} 0.143 (s)\\ \texttt{x 1.64} \end{tabular} & \begin{tabular}{c} 0.191 (s)\\ \texttt{x 2.19} \end{tabular} & \begin{tabular}{c} 0.545 (s)\\ \texttt{x 6.26} \end{tabular}\\ \hline
$40570$ & \begin{tabular}{c} 0.155 (s)\\ \texttt{x 1.00} \end{tabular} & \begin{tabular}{c} 0.461 (s)\\ \texttt{x 2.97} \end{tabular} & \begin{tabular}{c} 0.293 (s)\\ \texttt{x 1.89} \end{tabular} & \begin{tabular}{c} 0.377 (s)\\ \texttt{x 2.43} \end{tabular} & \begin{tabular}{c} 1.571 (s)\\ \texttt{x 10.12} \end{tabular}\\ \hline
$200424$ & \begin{tabular}{c} 1.014 (s)\\ \texttt{x 1.00} \end{tabular} & \begin{tabular}{c} 3.457 (s)\\ \texttt{x 3.41} \end{tabular} & \begin{tabular}{c} 1.911 (s)\\ \texttt{x 1.88} \end{tabular} & \begin{tabular}{c} 1.935 (s)\\ \texttt{x 1.91} \end{tabular} & \begin{tabular}{c} 8.742 (s)\\ \texttt{x 8.62} \end{tabular}\\ \hline
$580975$ & \begin{tabular}{c} 3.844 (s)\\ \texttt{x 1.00} \end{tabular} & \begin{tabular}{c} 9.767 (s)\\ \texttt{x 2.54} \end{tabular} & \begin{tabular}{c} 7.193 (s)\\ \texttt{x 1.87} \end{tabular} & \begin{tabular}{c} 6.804 (s)\\ \texttt{x 1.77} \end{tabular} & \begin{tabular}{c} 26.970 (s)\\ \texttt{x 7.02} \end{tabular}\\ \hline
$1747861$ & \begin{tabular}{c} 10.752 (s)\\ \texttt{x 1.00} \end{tabular} & \begin{tabular}{c} 31.203 (s)\\ \texttt{x 2.90} \end{tabular} & \begin{tabular}{c} 31.008 (s)\\ \texttt{x 2.88} \end{tabular} & \begin{tabular}{c} 26.069 (s)\\ \texttt{x 2.42} \end{tabular} & \begin{tabular}{c} 84.698 (s)\\ \texttt{x 7.88} \end{tabular}\\ \hline
\end{tabular}
\ifdefined\TabularWithCaption
}
\end{center}
\caption{Stiffness matrix (3D) : computational cost versus $n_{dof},$ with the \texttt{\OptVecS} Matlab/Octave/Python version ($2^{\mbox{nd}}/3^{\mbox{rd}}/4^{\mbox{th}}$ columns), with CXSparse ($1^{\mbox{st}}$ column) and FreeFEM++ ($5^{\mbox{th}}$ column) : time in seconds (top value) and speedup (bottom value). The speedup reference is CXSparse code.\label{table:CompStiffAssembling3DP1}}
\end{table}
\fi

\vspace{3mm}
\ifdefined\TabularWithCaption
\begin{table}[htbp]
\begin{center}
\noindent\adjustbox{max width=\textwidth}{
\fi
\begin{tabular}{@{}|r||*{5}{@{}c@{}|}@{}}
  \hline 
  $n_{dof}$ &  \begin{tabular}{c} CXSparse\\ (4.2.1) \end{tabular} &  \begin{tabular}{c} Matlab\\ (2014b) \end{tabular} &  \begin{tabular}{c} Octave\\ (3.8.1) \end{tabular} &  \begin{tabular}{c} Python\\ (3.4.0) \end{tabular} &  \begin{tabular}{c} FreeFEM\\ (3.31) \end{tabular}  \\ \hline \hline
$28444$ & \begin{tabular}{c} 0.023 (s)\\ \texttt{x 1.00} \end{tabular} & \begin{tabular}{c} 0.139 (s)\\ \texttt{x 5.92} \end{tabular} & \begin{tabular}{c} 0.115 (s)\\ \texttt{x 4.88} \end{tabular} & \begin{tabular}{c} 0.076 (s)\\ \texttt{x 3.25} \end{tabular} & \begin{tabular}{c} 0.762 (s)\\ \texttt{x 32.47} \end{tabular}\\ \hline
$111838$ & \begin{tabular}{c} 0.107 (s)\\ \texttt{x 1.00} \end{tabular} & \begin{tabular}{c} 0.412 (s)\\ \texttt{x 3.84} \end{tabular} & \begin{tabular}{c} 0.321 (s)\\ \texttt{x 2.99} \end{tabular} & \begin{tabular}{c} 0.288 (s)\\ \texttt{x 2.69} \end{tabular} & \begin{tabular}{c} 2.865 (s)\\ \texttt{x 26.75} \end{tabular}\\ \hline
$250020$ & \begin{tabular}{c} 0.267 (s)\\ \texttt{x 1.00} \end{tabular} & \begin{tabular}{c} 1.046 (s)\\ \texttt{x 3.92} \end{tabular} & \begin{tabular}{c} 0.727 (s)\\ \texttt{x 2.73} \end{tabular} & \begin{tabular}{c} 0.662 (s)\\ \texttt{x 2.48} \end{tabular} & \begin{tabular}{c} 6.432 (s)\\ \texttt{x 24.11} \end{tabular}\\ \hline
$1013412$ & \begin{tabular}{c} 1.133 (s)\\ \texttt{x 1.00} \end{tabular} & \begin{tabular}{c} 5.000 (s)\\ \texttt{x 4.41} \end{tabular} & \begin{tabular}{c} 4.238 (s)\\ \texttt{x 3.74} \end{tabular} & \begin{tabular}{c} 3.377 (s)\\ \texttt{x 2.98} \end{tabular} & \begin{tabular}{c} 26.301 (s)\\ \texttt{x 23.20} \end{tabular}\\ \hline
$2802258$ & \begin{tabular}{c} 3.142 (s)\\ \texttt{x 1.00} \end{tabular} & \begin{tabular}{c} 14.867 (s)\\ \texttt{x 4.73} \end{tabular} & \begin{tabular}{c} 11.483 (s)\\ \texttt{x 3.65} \end{tabular} & \begin{tabular}{c} 10.036 (s)\\ \texttt{x 3.19} \end{tabular} & \begin{tabular}{c} 72.561 (s)\\ \texttt{x 23.10} \end{tabular}\\ \hline
\end{tabular}
\ifdefined\TabularWithCaption
}
\end{center}
\caption{Elastic stiffness matrix (2D) : computational cost versus $n_{dof},$ with the \texttt{\OptVecS} Matlab/Octave/Python version ($2^{\mbox{nd}}/3^{\mbox{rd}}/4^{\mbox{th}}$ columns), with CXSparse ($1^{\mbox{st}}$ column) and FreeFEM++ ($5^{\mbox{th}}$ column) : time in seconds (top value) and speedup (bottom value). The speedup reference is CXSparse code.\label{table:CompStiffElasAssembling2DP1}}
\end{table}
\fi

\let\TabularWithCaption\undefined
 
\begin{table}[htbp]
\begin{center}
\noindent\adjustbox{max width=\textwidth}{\begin{tabular}{@{}c@{}}
\begin{tabular}{@{}cc@{}}
\textbf{StiffAssembling3DP1 - Matlab} & \textbf{StiffAssembling3DP1 - Octave} \\
\ifdefined\TabularWithCaption
\begin{table}[htbp]
\begin{center}
\noindent\adjustbox{max width=\textwidth}{
\fi
\begin{tabular}{@{}|r||*{5}{@{}c@{}|}@{}}
  \hline 
  $n_{dof}$ &  OptVS &  OptV &  OptV2 &  OptV1 &  base  \\ \hline \hline
$14708$ & \begin{tabular}{c} .135 {\tiny (s)}\\ \texttt{x 1} \end{tabular} & \begin{tabular}{c} .409 {\tiny (s)}\\ \texttt{x 3.03} \end{tabular} & \begin{tabular}{c} .442 {\tiny (s)}\\ \texttt{x 3.27} \end{tabular} & \begin{tabular}{c} 5.07 {\tiny (s)}\\ \texttt{x 37.5} \end{tabular} & \begin{tabular}{c} 25.6 {\tiny (s)}\\ \texttt{x 189} \end{tabular}\\ \hline
$40570$ & \begin{tabular}{c} .461 {\tiny (s)}\\ \texttt{x 1} \end{tabular} & \begin{tabular}{c} .827 {\tiny (s)}\\ \texttt{x 1.80} \end{tabular} & \begin{tabular}{c} .775 {\tiny (s)}\\ \texttt{x 1.68} \end{tabular} & \begin{tabular}{c} 14.0 {\tiny (s)}\\ \texttt{x 30.3} \end{tabular} & \begin{tabular}{c} 112 {\tiny (s)}\\ \texttt{x 244} \end{tabular}\\ \hline
$200424$ & \begin{tabular}{c} 3.46 {\tiny (s)}\\ \texttt{x 1} \end{tabular} & \begin{tabular}{c} 4.69 {\tiny (s)}\\ \texttt{x 1.36} \end{tabular} & \begin{tabular}{c} 4.74 {\tiny (s)}\\ \texttt{x 1.37} \end{tabular} & \begin{tabular}{c} 69.6 {\tiny (s)}\\ \texttt{x 20.1} \end{tabular} & \begin{tabular}{c} 3255 {\tiny (s)}\\ \texttt{x 942} \end{tabular}\\ \hline
$580975$ & \begin{tabular}{c} 9.77 {\tiny (s)}\\ \texttt{x 1} \end{tabular} & \begin{tabular}{c} 13.2 {\tiny (s)}\\ \texttt{x 1.35} \end{tabular} & \begin{tabular}{c} 14.1 {\tiny (s)}\\ \texttt{x 1.44} \end{tabular} & \begin{tabular}{c} 204 {\tiny (s)}\\ \texttt{x 20.9} \end{tabular} & -\\ \hline
$1747861$ & \begin{tabular}{c} 31.2 {\tiny (s)}\\ \texttt{x 1} \end{tabular} & \begin{tabular}{c} 40.4 {\tiny (s)}\\ \texttt{x 1.29} \end{tabular} & \begin{tabular}{c} 44.5 {\tiny (s)}\\ \texttt{x 1.42} \end{tabular} & \begin{tabular}{c} 623 {\tiny (s)}\\ \texttt{x 20.0} \end{tabular} & -\\ \hline
\end{tabular}
\ifdefined\TabularWithCaption
}
\end{center}
\caption{OptV methods for Matlab : Stiff matrix in 3d}
\end{table}
\fi

&
\ifdefined\TabularWithCaption
\begin{table}[htbp]
\begin{center}
\noindent\adjustbox{max width=\textwidth}{
\fi
\begin{tabular}{@{}|r||*{5}{@{}c@{}|}@{}}
  \hline 
  $n_{dof}$ &  OptVS &  OptV &  OptV2 &  OptV1 &  base  \\ \hline \hline
$14708$ & \begin{tabular}{c} .143 {\tiny (s)}\\ \texttt{x 1} \end{tabular} & \begin{tabular}{c} .177 {\tiny (s)}\\ \texttt{x 1.24} \end{tabular} & \begin{tabular}{c} .146 {\tiny (s)}\\ \texttt{x 1.02} \end{tabular} & \begin{tabular}{c} 76.0 {\tiny (s)}\\ \texttt{x 531} \end{tabular} & \begin{tabular}{c} 77.3 {\tiny (s)}\\ \texttt{x 540} \end{tabular}\\ \hline
$40570$ & \begin{tabular}{c} .293 {\tiny (s)}\\ \texttt{x 1} \end{tabular} & \begin{tabular}{c} .446 {\tiny (s)}\\ \texttt{x 1.52} \end{tabular} & \begin{tabular}{c} .488 {\tiny (s)}\\ \texttt{x 1.66} \end{tabular} & \begin{tabular}{c} 216 {\tiny (s)}\\ \texttt{x 737} \end{tabular} & \begin{tabular}{c} 248 {\tiny (s)}\\ \texttt{x 844} \end{tabular}\\ \hline
$200424$ & \begin{tabular}{c} 1.91 {\tiny (s)}\\ \texttt{x 1} \end{tabular} & \begin{tabular}{c} 2.37 {\tiny (s)}\\ \texttt{x 1.24} \end{tabular} & \begin{tabular}{c} 3.15 {\tiny (s)}\\ \texttt{x 1.65} \end{tabular} & \begin{tabular}{c} 1120 {\tiny (s)}\\ \texttt{x 586} \end{tabular} & \begin{tabular}{c} 3041 {\tiny (s)}\\ \texttt{x 1592} \end{tabular}\\ \hline
$580975$ & \begin{tabular}{c} 7.19 {\tiny (s)}\\ \texttt{x 1} \end{tabular} & \begin{tabular}{c} 9.15 {\tiny (s)}\\ \texttt{x 1.27} \end{tabular} & \begin{tabular}{c} 10.6 {\tiny (s)}\\ \texttt{x 1.47} \end{tabular} & \begin{tabular}{c} 3264 {\tiny (s)}\\ \texttt{x 454} \end{tabular} & -\\ \hline
$1747861$ & \begin{tabular}{c} 31.0 {\tiny (s)}\\ \texttt{x 1} \end{tabular} & \begin{tabular}{c} 38.0 {\tiny (s)}\\ \texttt{x 1.22} \end{tabular} & \begin{tabular}{c} 40.5 {\tiny (s)}\\ \texttt{x 1.31} \end{tabular} & - & -\\ \hline
\end{tabular}
\ifdefined\TabularWithCaption
}
\end{center}
\caption{OptV methods for Octave : Stiff matrix in 3d}
\end{table}
\fi

\end{tabular}
\vspace{0.15cm}\\

\begin{tabular}{@{}c@{}}
\textbf{StiffAssembling3DP1 - Python}\\
\ifdefined\TabularWithCaption
\begin{table}[htbp]
\begin{center}
\noindent\adjustbox{max width=\textwidth}{
\fi
\begin{tabular}{@{}|r||*{5}{@{}c@{}|}@{}}
  \hline 
  $n_{dof}$ &  OptVS &  OptV &  OptV2 &  OptV1 &  base  \\ \hline \hline
$14708$ & \begin{tabular}{c} .191 {\tiny (s)}\\ \texttt{x 1} \end{tabular} & \begin{tabular}{c} .207 {\tiny (s)}\\ \texttt{x 1.08} \end{tabular} & \begin{tabular}{c} .258 {\tiny (s)}\\ \texttt{x 1.35} \end{tabular} & \begin{tabular}{c} 12.8 {\tiny (s)}\\ \texttt{x 67.0} \end{tabular} & \begin{tabular}{c} 172 {\tiny (s)}\\ \texttt{x 903} \end{tabular}\\ \hline
$40570$ & \begin{tabular}{c} .377 {\tiny (s)}\\ \texttt{x 1} \end{tabular} & \begin{tabular}{c} .530 {\tiny (s)}\\ \texttt{x 1.41} \end{tabular} & \begin{tabular}{c} .823 {\tiny (s)}\\ \texttt{x 2.19} \end{tabular} & \begin{tabular}{c} 36.0 {\tiny (s)}\\ \texttt{x 95.5} \end{tabular} & \begin{tabular}{c} 488 {\tiny (s)}\\ \texttt{x 1295} \end{tabular}\\ \hline
$200424$ & \begin{tabular}{c} 1.93 {\tiny (s)}\\ \texttt{x 1} \end{tabular} & \begin{tabular}{c} 2.50 {\tiny (s)}\\ \texttt{x 1.29} \end{tabular} & \begin{tabular}{c} 4.15 {\tiny (s)}\\ \texttt{x 2.14} \end{tabular} & \begin{tabular}{c} 182 {\tiny (s)}\\ \texttt{x 94.0} \end{tabular} & \begin{tabular}{c} 2480 {\tiny (s)}\\ \texttt{x 1282} \end{tabular}\\ \hline
$580975$ & \begin{tabular}{c} 6.80 {\tiny (s)}\\ \texttt{x 1} \end{tabular} & \begin{tabular}{c} 8.89 {\tiny (s)}\\ \texttt{x 1.31} \end{tabular} & \begin{tabular}{c} 12.4 {\tiny (s)}\\ \texttt{x 1.83} \end{tabular} & \begin{tabular}{c} 541 {\tiny (s)}\\ \texttt{x 79.5} \end{tabular} & -\\ \hline
$1747861$ & \begin{tabular}{c} 26.1 {\tiny (s)}\\ \texttt{x 1} \end{tabular} & \begin{tabular}{c} 34.2 {\tiny (s)}\\ \texttt{x 1.31} \end{tabular} & \begin{tabular}{c} 40.3 {\tiny (s)}\\ \texttt{x 1.55} \end{tabular} & - & -\\ \hline
\end{tabular}
\ifdefined\TabularWithCaption
}
\end{center}
\caption{OptV methods for Python : Stiff matrix in 3d}
\end{table}
\fi

\end{tabular}
\end{tabular}}
\vspace{0.15cm}

\captionof{table}{Stiffness matrix (3D) : comparison of \texttt{\OptVecS}, \texttt{\OptVec}, \texttt{OptV2}, \texttt{OptV1} and \texttt{base} codes in Matlab (top left),
              Octave (top right) and Python (bottom) giving
              time in seconds (top value) and \texttt{\OptVecS} speedup (bottom value).}\label{tab:Stiff3D_Merge}
\end{center}
\end{table}

\begin{table}[htbp]
\begin{center}
\noindent\adjustbox{max width=\textwidth}{\begin{tabular}{@{}c@{}}
\begin{tabular}{@{}cc@{}}
\textbf{StiffElasAssembling2DP1 - Matlab} & \textbf{StiffElasAssembling2DP1 - Octave} \\
\ifdefined\TabularWithCaption
\begin{table}[htbp]
\begin{center}
\noindent\adjustbox{max width=\textwidth}{
\fi
\begin{tabular}{@{}|r||*{5}{@{}c@{}|}@{}}
  \hline 
  $n_{dof}$ &  OptVS &  OptV &  OptV2 &  OptV1 &  base  \\ \hline \hline
$28444$ & \begin{tabular}{c} .114 {\tiny (s)}\\ \texttt{x 1} \end{tabular} & \begin{tabular}{c} .272 {\tiny (s)}\\ \texttt{x 2.39} \end{tabular} & \begin{tabular}{c} .185 {\tiny (s)}\\ \texttt{x 1.63} \end{tabular} & \begin{tabular}{c} 16.9 {\tiny (s)}\\ \texttt{x 148} \end{tabular} & \begin{tabular}{c} 67.8 {\tiny (s)}\\ \texttt{x 594} \end{tabular}\\ \hline
$250020$ & \begin{tabular}{c} 1.01 {\tiny (s)}\\ \texttt{x 1} \end{tabular} & \begin{tabular}{c} 1.68 {\tiny (s)}\\ \texttt{x 1.65} \end{tabular} & \begin{tabular}{c} 1.91 {\tiny (s)}\\ \texttt{x 1.88} \end{tabular} & \begin{tabular}{c} 150 {\tiny (s)}\\ \texttt{x 148} \end{tabular} & \begin{tabular}{c} 6156 {\tiny (s)}\\ \texttt{x 6073} \end{tabular}\\ \hline
$686164$ & \begin{tabular}{c} 3.20 {\tiny (s)}\\ \texttt{x 1} \end{tabular} & \begin{tabular}{c} 5.28 {\tiny (s)}\\ \texttt{x 1.65} \end{tabular} & \begin{tabular}{c} 5.42 {\tiny (s)}\\ \texttt{x 1.70} \end{tabular} & \begin{tabular}{c} 414 {\tiny (s)}\\ \texttt{x 129} \end{tabular} & -\\ \hline
$1771042$ & \begin{tabular}{c} 8.97 {\tiny (s)}\\ \texttt{x 1} \end{tabular} & \begin{tabular}{c} 14.7 {\tiny (s)}\\ \texttt{x 1.64} \end{tabular} & \begin{tabular}{c} 15.2 {\tiny (s)}\\ \texttt{x 1.69} \end{tabular} & \begin{tabular}{c} 1090 {\tiny (s)}\\ \texttt{x 122} \end{tabular} & -\\ \hline
$3957204$ & \begin{tabular}{c} 21.6 {\tiny (s)}\\ \texttt{x 1} \end{tabular} & \begin{tabular}{c} 34.1 {\tiny (s)}\\ \texttt{x 1.58} \end{tabular} & \begin{tabular}{c} 33.1 {\tiny (s)}\\ \texttt{x 1.53} \end{tabular} & - & -\\ \hline
\end{tabular}
\ifdefined\TabularWithCaption
}
\end{center}
\caption{OptV methods for Matlab : StiffElas matrix in 2d}
\end{table}
\fi

&
\ifdefined\TabularWithCaption
\begin{table}[htbp]
\begin{center}
\noindent\adjustbox{max width=\textwidth}{
\fi
\begin{tabular}{@{}|r||*{5}{@{}c@{}|}@{}}
  \hline 
  $n_{dof}$ &  OptVS &  OptV &  OptV2 &  OptV1 &  base  \\ \hline \hline
$28444$ & \begin{tabular}{c} .082 {\tiny (s)}\\ \texttt{x 1} \end{tabular} & \begin{tabular}{c} .129 {\tiny (s)}\\ \texttt{x 1.57} \end{tabular} & \begin{tabular}{c} .091 {\tiny (s)}\\ \texttt{x 1.11} \end{tabular} & \begin{tabular}{c} 63.7 {\tiny (s)}\\ \texttt{x 777} \end{tabular} & \begin{tabular}{c} 88.6 {\tiny (s)}\\ \texttt{x 1080} \end{tabular}\\ \hline
$250020$ & \begin{tabular}{c} .726 {\tiny (s)}\\ \texttt{x 1} \end{tabular} & \begin{tabular}{c} 1.26 {\tiny (s)}\\ \texttt{x 1.74} \end{tabular} & \begin{tabular}{c} .915 {\tiny (s)}\\ \texttt{x 1.26} \end{tabular} & \begin{tabular}{c} 564 {\tiny (s)}\\ \texttt{x 777} \end{tabular} & \begin{tabular}{c} 4485 {\tiny (s)}\\ \texttt{x 6176} \end{tabular}\\ \hline
$686164$ & \begin{tabular}{c} 2.21 {\tiny (s)}\\ \texttt{x 1} \end{tabular} & \begin{tabular}{c} 3.57 {\tiny (s)}\\ \texttt{x 1.61} \end{tabular} & \begin{tabular}{c} 3.36 {\tiny (s)}\\ \texttt{x 1.52} \end{tabular} & \begin{tabular}{c} 1550 {\tiny (s)}\\ \texttt{x 701} \end{tabular} & -\\ \hline
$1771042$ & \begin{tabular}{c} 6.85 {\tiny (s)}\\ \texttt{x 1} \end{tabular} & \begin{tabular}{c} 10.7 {\tiny (s)}\\ \texttt{x 1.56} \end{tabular} & \begin{tabular}{c} 9.39 {\tiny (s)}\\ \texttt{x 1.37} \end{tabular} & - & -\\ \hline
$3957204$ & \begin{tabular}{c} 16.2 {\tiny (s)}\\ \texttt{x 1} \end{tabular} & \begin{tabular}{c} 25.8 {\tiny (s)}\\ \texttt{x 1.59} \end{tabular} & \begin{tabular}{c} 22.0 {\tiny (s)}\\ \texttt{x 1.36} \end{tabular} & - & -\\ \hline
\end{tabular}
\ifdefined\TabularWithCaption
}
\end{center}
\caption{OptV methods for Octave : StiffElas matrix in 2d}
\end{table}
\fi

\end{tabular}
\vspace{0.15cm}\\

\begin{tabular}{@{}c@{}}
\textbf{StiffElasAssembling2DP1 - Python}\\
\ifdefined\TabularWithCaption
\begin{table}[htbp]
\begin{center}
\noindent\adjustbox{max width=\textwidth}{
\fi
\begin{tabular}{@{}|r||*{5}{@{}c@{}|}@{}}
  \hline 
  $n_{dof}$ &  OptVS &  OptV &  OptV2 &  OptV1 &  base  \\ \hline \hline
$28444$ & \begin{tabular}{c} .136 {\tiny (s)}\\ \texttt{x 1} \end{tabular} & \begin{tabular}{c} .207 {\tiny (s)}\\ \texttt{x 1.53} \end{tabular} & \begin{tabular}{c} .202 {\tiny (s)}\\ \texttt{x 1.49} \end{tabular} & \begin{tabular}{c} 30.0 {\tiny (s)}\\ \texttt{x 221} \end{tabular} & \begin{tabular}{c} 183 {\tiny (s)}\\ \texttt{x 1348} \end{tabular}\\ \hline
$250020$ & \begin{tabular}{c} .721 {\tiny (s)}\\ \texttt{x 1} \end{tabular} & \begin{tabular}{c} 1.08 {\tiny (s)}\\ \texttt{x 1.50} \end{tabular} & \begin{tabular}{c} 1.22 {\tiny (s)}\\ \texttt{x 1.7} \end{tabular} & \begin{tabular}{c} 277 {\tiny (s)}\\ \texttt{x 384} \end{tabular} & \begin{tabular}{c} 1639 {\tiny (s)}\\ \texttt{x 2274} \end{tabular}\\ \hline
$686164$ & \begin{tabular}{c} 2.05 {\tiny (s)}\\ \texttt{x 1} \end{tabular} & \begin{tabular}{c} 3.06 {\tiny (s)}\\ \texttt{x 1.50} \end{tabular} & \begin{tabular}{c} 3.55 {\tiny (s)}\\ \texttt{x 1.73} \end{tabular} & \begin{tabular}{c} 761 {\tiny (s)}\\ \texttt{x 372} \end{tabular} & -\\ \hline
$1771042$ & \begin{tabular}{c} 6.06 {\tiny (s)}\\ \texttt{x 1} \end{tabular} & \begin{tabular}{c} 9.12 {\tiny (s)}\\ \texttt{x 1.50} \end{tabular} & \begin{tabular}{c} 9.58 {\tiny (s)}\\ \texttt{x 1.58} \end{tabular} & - & -\\ \hline
$3957204$ & \begin{tabular}{c} 14.0 {\tiny (s)}\\ \texttt{x 1} \end{tabular} & \begin{tabular}{c} 21.2 {\tiny (s)}\\ \texttt{x 1.52} \end{tabular} & \begin{tabular}{c} 21.5 {\tiny (s)}\\ \texttt{x 1.54} \end{tabular} & - & -\\ \hline
\end{tabular}
\ifdefined\TabularWithCaption
}
\end{center}
\caption{OptV methods for Python : StiffElas matrix in 2d}
\end{table}
\fi

\end{tabular}
\end{tabular}}
\vspace{0.15cm}

\captionof{table}{Elastic stiffness matrix (2D): comparison of \texttt{\OptVecS}, \texttt{\OptVec}, \texttt{OptV2} , \texttt{OptV1} and \texttt{base} codes in Matlab (top left),
              Octave (top right) and Python (bottom) giving
              time in seconds (top value) and \texttt{\OptVecS} speedup (bottom value).}\label{tab:StiffElas2D_Merge}
\end{center}
\end{table}

\section{Proof of Lemma~\ref{lem:VecLocalElas}}
\label{app:ProofLem}
To prove Lemma~\ref{lem:VecLocalElas}, we introduce the following matrix $\MAT{B}_l$ :
$$
\MAT{B}_l=\begin{pmatrix}
\delta_{l,1} & 0\\ 0 & \delta_{l,2}\\ \delta_{l,2} & \delta_{l,1}
\end{pmatrix}\ \ \mbox{if}\ d=2,\  \
\mbox{and}\ \
\MAT{B}_l=\begin{pmatrix}
\delta_{l,1} & 0 & 0 \\ 0 & \delta_{l,2} & 0\\ 0 & 0 & \delta_{l,3}\\
\delta_{l,2} & \delta_{l,1} & 0\\
0 & \delta_{l,3} & \delta_{l,2}\\
\delta_{l,3}& 0 & \delta_{l,1}
\end{pmatrix}\ \ \mbox{if}\ d=3.
$$
Thus we have $\Odv(\BaryCoorVec_{l,\il})=\MAT{B}_l \GRAD\BaryCoor_\il$ and then
$$
\Odv^t(\BaryCoorVec_{n,\jl}) \MAT{C}\Odv(\BaryCoorVec_{l,\il})
=\GRAD\BaryCoor_\jl^t \MAT{B}_n^t\MAT{C}\MAT{B}_l \GRAD\BaryCoor_\il.
$$
Moreover we have $\MAT{C}=\lambda\MAT{C}_0+\mu\MAT{C}_1$ with
$$\mathbb{C}_0=
\begin{pmatrix}
  \mathds{1}_d&   \mathds{O}_{d,2d-3}\\
 \mathds{O}_{2d-3,d} &  \mathds{O}_{2d-3}
\end{pmatrix}_{3(d-1)\times 3(d-1)}
\ \ \mbox{and}\ \
\mathbb{C}_1=
\begin{pmatrix}
  2\mathbb{I}_d&   \mathds{O}_{d,2d-3}\\
 \mathds{O}_{2d-3,d} &  \mathbb{I}_{2d-3}
\end{pmatrix}_{3(d-1)\times 3(d-1)}
$$
Thus we obtain
\begin{equation*}
\Odv^t(\BaryCoorVec_{n,\jl}) \MAT{C}\Odv(\BaryCoorVec_{l,\il})=
\lambda \GRAD\BaryCoor_\jl^t \MAT{B}_n^t\MAT{C}_0\MAT{B}_l \GRAD\BaryCoor_\il
+ \mu \GRAD\BaryCoor_\jl^t \MAT{B}_n^t\MAT{C}_1\MAT{B}_l \GRAD\BaryCoor_\il.
\end{equation*}
Denoting $\MAT{Q}^{n,l}=\MAT{B}_n^t\MAT{C}_0\MAT{B}_l$ and $\MAT{S}^{n,l}=\MAT{B}_n^t\MAT{C}_1\MAT{B}_l$
we obtain~\eqref{eq:VecLocalElas} which ends the proof of Lemma~\ref{lem:VecLocalElas}.

\section{Remaining routines}
\label{Appendix:Gradients}

\subsection{Gradients of the barycentric coordinates  \label{subsec:gradients}}
Let $T_k$ be a $d$-simplex of $\R^d$ with vertices $\q^0,\hdots,\q^d$, and
$\hat{T}$ be the reference $d$-simplex with vertices $\hat{\q}^0,\hdots,\hat{\q}^d$ where
$\hat{\q}^0=\vecb{0}_d$ and $\hat{\q}^i=\vecb{e}_i,$ $\forall i\in \ENS{1}{d}.$

Let $\mathcal{F}_k$ be the bijection from $\hat{T}$ to $T_k$ defined by
$\q=\mathcal{F}_k(\hat{\q})=\MAT{B}_k \hat{\q} +\q^0$
where $\MAT{B}_k \in \MS{d}{\R}$ is such that its $i$-th column is equal to $\q^i -\q^0,$ for all $i\in\ENS{1}{d}.$

The barycentric coordinates of $\hat{\q}=(\hat{x}_1,\hdots,\hat{x}_d)\in \hat{T}$ are given by
$\hat{\lambda}_0=1-\sum_{i=1}^d\hat{x}_i,$ and $\hat{\lambda}_{i}=\hat{x}_i,$ $\forall i\in \ENS{1}{d}.$
The barycentric coordinates of $\q=(x_1,\hdots,x_d)\in T_k$ are given by
$\lambda_{k,i}(\q)=\hat{\lambda}_i\circ \mathcal{F}^{-1}_k(\q)$ and we have
\begin{equation}\label{gradient:systems}
\GRAD\lambda_{k,i}(\q) = \MAT{B}_k^{-t} \hat{\GRAD}\hat{\lambda}_i(\hat{\q}),\ \forall i\in \ENS{0}{d},
\end{equation}
with
$\hat{\GRAD}\hat{\lambda}_0(\hat{\q})=\begin{pmatrix} -1\\\hdots \\-1\end{pmatrix},$ $\hat{\GRAD}\hat{\lambda}_i=\vecb{e}_i,$
$\forall i\in \ENS{1}{d}.$ Note that gradients are constant.
Let
$$\hat{\MAT{G}}=\begin{pmatrix} \hat{\GRAD}\hat{\lambda}_0, &\hdots, & \hat{\GRAD}\hat{\lambda}_d\end{pmatrix}=\begin{pmatrix}
-1 & 1 & 0 & \hdots   &0 \\
-1 & 0 & 1 &  \ddots & \vdots \\
\vdots & \vdots & \ddots & \ddots  & 0\\
-1 & 0 & \hdots & 0 & 1
\end{pmatrix}.$$
Then computing the gradients of the barycentric coordinates is equivalent to solve~$(d+1)$ linear systems, written
in matrix form as follows:
\begin{equation}\label{gradient:SystemeMat}
\MAT{B}_k^{t} \MAT{G}_k = \hat{\MAT{G}},
\end{equation}
where $\MAT{G}_k=\begin{pmatrix}\GRAD\lambda_{k,0}(\q),& \hdots ,& \GRAD\lambda_{k,d}(\q)\end{pmatrix}\in\MS{d,d+1}{\R}.$

For each $d$-simplex one has to calculate $(d+1)$ gradients and thus to determine
$(d+1)\nme$ vectors of dimension $d$.

The vectorization of the calculation of the gradients is done by rewriting
the equations \eqref{gradient:SystemeMat}, for $k=1,...,\nme$, under an equivalent form
of a large block diagonal sparse system of size $N=d\times\nme$, with $d$-by-$d$ diagonal blocks given by:
\begin{equation}
\begin{pmatrix}
\MAT{B}_1^{t} & \MAT{O} & \hdots & \MAT{O}\\
 \MAT{O} & \ddots & \ddots & \vdots\\
 \vdots &  \ddots & \ddots & \MAT{O} \\
 \MAT{O}& \hdots & \MAT{O}&\MAT{B}_{\nme}^{t}
\end{pmatrix}_{N\times N}
\begin{pmatrix}
\MAT{G}_1\\
\MAT{G}_2\\
\vdots\\
\MAT{G}_{\nme}\\
\end{pmatrix}_{N\times(d+1)}=
\begin{pmatrix}
\hat{\MAT{G}}\\
\hat{\MAT{G}}\\
\vdots\\
\hat{\MAT{G}}\\
\end{pmatrix}_{N\times(d+1)}
\end{equation}

\begin{algorithm}[H]
\captionsetup{font=footnotesize}
\caption{Vectorized computation of gradients of the basis functions in dimension $d$}
\label{algo:GradientVec}
\ifdefined\IsInputOutput
\Input{14.cm}{
  \begin{tabular}[t]{lcl}
    $\q$   & : & $d$-by-$\nq$ array \\
    $\me$  & : & $(d+1)$-by-$\nme$ connectivity array \\
  \end{tabular}
}
\Output{14.cm}{
  \begin{tabular}{lcl}
    $\MAT{G}$ & : & array of the gradients ($\nme$-by-$(d+1)$-by-$d$)\\
    & & $\MAT{G}(k,\il,:)=\GRAD\BasisFunc_{\il}^k(\q),$ $\forall \il\in\ENS{1}{d+1}$
  \end{tabular}
}
\hrule
\fi
\begin{algorithmic}[0]
\Function{$\footnotesize \vecb{G}\gets \FNameDef{GradientVec}$}{$\q,\me$}
\State $\MAT{K} \gets \MAT{I} \gets \MAT{J} \gets \FNameStd{zeros}(d,d,\nme)$
\State $\Var{ii} \gets d*[0:(\nme-1)]$
\For{i}{1}{d}
  \For{j}{1}{d}
    \State $\MAT{K}(i,j,:) \gets \q(i,\me(j+1,:)) - \q(i,\me(1,:))$
    \State $\MAT{I}(i,j,:) \gets \Var{ii}+j,\ \MAT{J}(i,j,:) \gets \Var{ii}+i$
  \EndFor
\EndFor
\State $\MAT{S} \gets \FNameStd{sparse}(\MAT{I}(:),\MAT{J}(:),\MAT{K}(:),d*\nme,d*\nme)$
\State $\MAT{R} \gets \FNameStd{zeros}(d*\nme,d+1)$ \Comment{Build RHS}
\State $\hat{\MAT{G}} \gets [-\mathds{1}_{d\times 1},\MAT{I}_{d}]$
\State $\MAT{R} \gets \FNameStd{copymat}(\hat{\MAT{G}},\nme,1)$
\State $\MAT{G} \gets \FNameStd{solve}(\MAT{S},\MAT{R})$ \Comment{$\MAT{G}(d(k-1)+i,\il)=\DP{\BaCo_\il}{x_i}_{|T_k}$}
\State $\MAT{G} \gets \FNameStd{transform}(\MAT{G},...)$ \Comment{such that $\MAT{G}(k,\il,i)=\DP{\BaCo_\il}{x_i}_{|T_k}$}
\EndFunction
\end{algorithmic}
\end{algorithm}

The performance of this algorithm may be improved by writing specific algorithms in each dimension $d=1,2$ or $3$
(see Appendix~A in~\cite{CJS:RR:2014}).

\subsection{Elastic stiffness matrix assembly : algorithm using the symmetry}\label{Appendix:StiffElasOptV4S}
When the assembly matrix is symmetric, one may improve the performance of
Algorithm~\ref{algo:AssemblyGenP1vectorOptV4} by using the symmetry of the element matrices (see
Section 4), which leads to the following algorithm:
\begin{algorithm}[H]
\captionsetup{font=footnotesize}
\caption{(\texttt{\OptVecS}) - Optimized assembly in vector case ($m>1$)}
\label{algo:AssemblyGenP1vectorOptV4s}
\ifdefined\IsInputOutput
\begin{minipage}{14.cm}
\footnotesize{\textbf{Input :}\\
  \begin{tabular}[t]{lcl}
     $\me$&:& $\ndfe$-by-$\nme$ connectivity array \\
          &  &with $\ndfe=d+1$\\
  \end{tabular}\\
\textbf{Output :}\\
  \begin{tabular}{lcl}
    $\MAT{M}$ & : & $\ndof$-by-$\ndof$ sparse matrix \\
                & &where $\ndof=m \nq.$
  \end{tabular}\\
}
\end{minipage}
\fi
\begin{algorithmic}[1]
\Function{$\MAT{M} \gets $\FNameDef{AssemblyVecGenP1\OptVecS}}{$\me,\nq,\hdots$}
\State $\ndof \gets m*\nq$ 
\State $\MAT{M} \gets \MAT{O}_{\ndof}$ \Comment{$\ndof$-by-$\ndof$ sparse matrix}
\For{l}{1}{m}
  \For{\il}{1}{d+1}
    \State $ \vecb{I}_g \gets  m*(\me(\il,:)-1)+l$
    \State $ii \gets m(\il-1)+l$
    \For{n}{1}{m}
       \For{\jl}{1}{d+1}
          \State $jj \gets m(\jl-1)+n$
          \If{$ii>jj$}
	  \State $ \vecb{K}_g \gets \FName{vecHe}(l,\il,n,\jl,\hdots)$
	  \State $ \vecb{J}_g \gets  m*(\me(\jl,:)-1)+n$
	  \State $\MAT{M} \gets \MAT{M} + \FNameStd{Sparse}(\vecb{I}_g,\vecb{J}_g,\vecb{K}_g,\ndof,\ndof)$
	  \EndIf
      \EndFor
    \EndFor
  \EndFor
\EndFor
\State $\MAT{M} \gets \MAT{M} + \MAT{M}' $
\For{l}{1}{m}
  \For{\il}{1}{d+1}
    \State $ \vecb{I}_g \gets  m*(\me(\il,:)-1)+l$
    \State $ \vecb{K}_g \gets  \FName{vecHe}(l,\il,l,\il,\hdots)$
    \State $\MAT{M} \gets \MAT{M} + \FNameStd{Sparse}(\vecb{I}_g,\vecb{I}_g,\vecb{K}_g,\ndof,\ndof)$
  \EndFor
\EndFor
\EndFunction
\end{algorithmic}
\end{algorithm}

\section{Extension to $\Pk{k}$-Lagrange finite elements\label{sec:PkFEM}}

In this section we adapt the optimized algorithm of Section~\ref{sec:CodeOptV2}
to the case of finite elements of higher order.
For simplicity, we consider the assembly algorithm on the example of the mass matrix.

The mesh used is adapted to $\Pk{k}$ finite elements and is called a``$\Pk{k}$-mesh''.
Only arrays $\q$ and~$\me$ differ between the usual mesh and the $\Pk{k}$-mesh.
In the $\Pk{k}$-mesh, $\q$ contains the coordinates of the nodal points associated to the $\Pk{k}$
finite elements and $\me$ is of dimension
$\ndfe$-by-$\nme$, where $\ndfe$ is the local number of $\Pk{k}$-nodes in a $d$-simplex~$K$ :
$\ndfe=\frac{(d+k)!}{d!k!}$, as shown in the table below.
\begin{center}
\begin{tabular}{lccl}
\hline
\textbf{name} & \textbf{type} & \textbf{dimension} & \textbf{description}\\
\hline
$\ndfe$ & integer & 1 & local number of $\Pk{k}$-nodes in a $d$-simplex\\
$\nq$ & integer & 1 & number of $\Pk{k}$-nodes\\
$\q$   & double  &$\dd\times \nq$ &
\begin{minipage}[t]{5.9cm}
array of $\Pk{k}$-node coordinates
\end{minipage}\\
$\me$  & integer & $\ndfe\times \nme$ &
\begin{minipage}[t]{5.9cm}
($\Pk{k}$) connectivity array
\end{minipage}\\
\hline
\end{tabular}
\end{center}

By construction, the total number of degrees of freedom of a $\Pk{k}$-mesh
is its number of nodal points.
One may use for example \texttt{gmsh}~\cite{gmsh:2014} to generate a $\Pk{k}$-mesh in 2D or in 3D.

First, we need to introduce some notations: let $\SMIE_{d}^k$ be the set of multi-indices given by
\begin{equation}
\SMIE_{d}^k=\left\{
\mi{\alpha}=(\alpha_1,\hdots,\alpha_{d+1})\in \N^{d+1} \ \mbox{such that}\ |\mi{\alpha}|:= \sum_{i=1}^{d+1} \alpha_i  =k
\right\},
\end{equation}
with $\#\SMIE_{d}=N$.
Then the $\mathbb{P}_k$ basis functions $\PkBasisFunc_{\mi{\alpha}}$ on a $d$-simplex $K$
may be deduced from the barycentric coordinates $\{\BaCo_j\}_{j=1}^{d+1}$
\begin{equation}\label{Pk:functions}
\PkBasisFunc_{\mi{\alpha}}=\prod_{l=1}^{d+1} \prod_{j=0}^{\alpha_l-1} \frac{k\lambda_l-j}{j+1},\
\ \ \forall \mi{\alpha}\in \SMIE_d^k,
\end{equation}
or equivalently, noticing that $\PkBasisFunc_{\mi{\alpha}}$ is
a polynomial in the variable $(\BaCo_1,\hdots,\BaCo_{d+1})$ and introducing a  multi-index
$\mi{\mu}=(\mu_1,\hdots,\mu_{d+1})\in \N^{d+1}$, we have
\begin{align}
\PkBasisFunc_{\mi{\alpha}}&=
\sum_{|\mi{\mu}| \leq k} a_{\mi{\mu}}(\mi{\alpha}) \left(\prod_{j=1}^{d+1} \BaCo_j^{\mu_j}\right).
\label{PolynomialRepresentation}
\end{align}
All the non-zero $a_{\mi{\mu}}(\mi{\alpha})$ values can be computed from \eqref{Pk:functions}
and depend only on $\mi{\alpha},$ $d$ and $k$.

As in the previous sections, the assembly algorithm of the mass matrix is based on the vectorization
of the local mass matrix $\Mass^e$ on $K$, which is an $N$-by-$N$ matrix given by

$$\Mass_{I(\mi{\alpha}),I(\mi{\beta})}^e(K)=\int_K \PkBasisFunc_{\mi{\alpha}} \PkBasisFunc_{\mi{\beta}} d\q,  \ \ \forall (\mi{\alpha},\mi{\beta})\in \SMIE_d^k\times \SMIE_d^k,$$
where $\foncdefsmall{I}{\SMIE_d^k}{\ENS{1}{N}}$ is the local numbering choice.

We then introduce a formula of the same type as~(\ref{eq:MatElemMassWP1}) to vectorize the computation
of $\Mass^e$.
Using \eqref{PolynomialRepresentation}, we have for all $(\mi{\alpha},\mi{\beta})\in \SMIE_d^k\times \SMIE_d^k$
\begin{align*}
\int_K  \PkBasisFunc_{\mi{\alpha}} \PkBasisFunc_{\mi{\beta}}d\q
= \sum_{ |\mi{\mu}| \leq k}
\sum_{ |\mi{\nu}| \leq k} a_{\mi{\mu}}(\mi{\alpha}) a_{\mi{\nu}}(\mi{\beta})
\int_K  \prod_{j=1}^{d+1} \BaCo_j^{\mu_j+\nu_j} d\q.
\end{align*}
Then, using formula \eqref{MagicFormula} we obtain
\begin{equation}\label{eq:MatElemMassPk}
\int_K  \PkBasisFunc_{\mi{\alpha}} \PkBasisFunc_{\mi{\beta}}d\q
= d!|K| C_{\mi{\alpha},\mi{\beta}},
\end{equation}
where the constant $C_{\mi{\alpha},\mi{\beta}}$ does not depend on $K$ and is given by
\begin{equation}
C_{\mi{\alpha},\mi{\beta}}=\sum_{ |\mi{\mu}| \leq k}
\sum_{|\mi{\nu}| \leq k} a_{\mi{\mu}}(\mi{\alpha}) a_{\mi{\nu}}(\mi{\beta})
\frac{\prod\limits_{i=1}^{d+1} (\mu_i+\nu_i)! }{(d+|\mi{\mu}|+|\mi{\nu}|  )!}.
\end{equation}

Using~\eqref{eq:MatElemMassPk}, we can now extend Algorithm~\ref{algo:AssemblyMassWP1} (with $w=1$)
to the $\Pk{k}$ finite element case.
This leads to the vectorized algorithm of the mass matrix given in Algorithm~\ref{algo:AssemblyMassPk}.

\begin{remark}
We have considered the extension of the \texttt{OptV2} algorithm to finite elements of higher order.
The main idea is that all the steps of Section~\ref{sec:CodeOptV2} remain valid for $\Pk{k}$ finite elements,
if one replaces $(d+1)$ by $\ndfe$, and with $\q$ and $\me$ defined above. Then, one may derive from
Algorithm~\ref{algo:AssemblyMassPk} the other optimized versions \texttt{\OptVec} and \texttt{\OptVecS}
for the $\Pk{k}$ case, as in Section~\ref{sec:CodeOptV2}.
\end{remark}

\begin{algorithm}[H]
\captionsetup{font=footnotesize}
\caption{(\texttt{OptV2}) - Mass matrix in $\Pk{k}$ case }
\label{algo:AssemblyMassPk}
\ifdefined\IsInputOutput
\begin{minipage}{14.cm}
\footnotesize{\textbf{Input :}\\
  \begin{tabular}[t]{lcl}
     $\me$&:& $\ndfe$-by-$\nme$ connectivity array with $\ndfe=\frac{(d+k)!}{d!k!}$\\
  \end{tabular}\\
\textbf{Output :}\\
  \begin{tabular}{lcl}
    $\MAT{M}$ & : & $\nq$-by-$\nq$ sparse matrix
  \end{tabular}\\
}
\end{minipage}
\fi
\begin{algorithmic}[1]
\Function{$\MAT{M} \gets $\FNameDef{AssemblyMassPk}}{$\me,\volumes,\nq,d,k$}
\State $\MAT{C} \gets \FName{coeffMass}(d,k)$ \Comment{Get coefficients $C_{\mi{\alpha},\mi{\beta}}$}
\State $\Kg \gets \Ig \gets \Jg \gets \FNameStd{zeros}(\ndfes,\nme)$\Comment{$\ndfes$-by-$\nme$ 2d-arrays}
\State $l \gets 1$
\For{\jl}{1}{\ndfe}
  \For{\il}{1}{\ndfe}
   \State $ \Kg(l,:) \gets d! *\MAT{C}(\il,\jl) * \volumes$
   \State $ \Ig(l,:) \gets  \me(\il,:)$
   \State $ \Jg(l,:) \gets  \me(\jl,:)$
   \State $ l \gets l+1$
  \EndFor
\EndFor
\State $\MAT{M} \gets \FNameStd{Sparse}(\Ig,\Jg,\Kg,\nq,\nq)$
\EndFunction
\end{algorithmic}
\end{algorithm}

In Table~\ref{MassPk}, using Matlab, we show the computation times versus the number of $\Pk{k}$ nodes,
for Algorithm~\ref{algo:AssemblyMassWP1} (with $w=1$), and for Algorithm~\ref{algo:AssemblyMassPk}
with $k=1,2,3,4,5,6$. We observe that the computation times are almost the same for
Algorithm~\ref{algo:AssemblyMassWP1} and Algorithm~\ref{algo:AssemblyMassPk} with $k=1$.
Moreover, for a fixed number of nodes, the computation times increase slowly with the degree
of the polynomials: for a million of nodes, the computation time with $\Pk{5}$ finite elements is twice
the one for $\Pk{1}$ finite elements.

\def\TabularWithCaption{}
\ifdefined\TabularWithCaption
\begin{table}[htbp]
\begin{center}
\noindent\adjustbox{max width=\textwidth}{
\fi
\begin{tabular}{|r||*{7}{c|}}
  \hline 
  $\ndof$ &  P1OptV2 &  Pk(k=1) &  Pk(k=2) &  Pk(k=3) &  Pk(k=4) &  Pk(k=5) &  Pk(k=6)  \\ \hline \hline
$3.10^4$ & 0.535 & 0.543 & 0.459 & 0.544 & 0.707 & 0.982 & 1.350\\ \hline
$1.2 10^5$ & 2.322 & 2.500 & 2.025 & 2.407 & 3.184 & 4.389 & 5.696\\ \hline
$5. 10^5$ & 10.885 & 13.203 & 9.811 & 11.684 & 15.184 & 19.766 & 25.340\\ \hline
$10^6$ & 22.744 & 28.362 & 22.635 & 25.656 & 33.314 & 42.812 & 54.782\\ \hline
\end{tabular}
\ifdefined\TabularWithCaption
}
\end{center}
\caption{3D Mass matrix : computational cost versus $\ndof$,
with Matlab, for \texttt{OptV2} code : Algorithm~\ref{algo:AssemblyMassWP1} with $w=1$) (column $1$),
and with Algorithm~\ref{algo:AssemblyMassPk}
for $k=1,2,3,4,5,6$ (columns $2$ to $7$).\label{MassPk}}
\end{table}
\fi

\begin{acknowledgements}
The authors would like to thank Prof. H-P. Langtangen for his many constructive
comments that led to a better presentation of the paper.
\end{acknowledgements}

\bibliography{paperBITrevised}
\bibliographystyle{plain}

\end{document}